\documentclass[12pt,epsf]{article}
\usepackage{epsf}
\usepackage{graphicx,graphics,cite}
\usepackage{subfigure}
\usepackage[dvipsnames]{xcolor}

\usepackage{graphicx,epsfig}
\usepackage{simplewick}
\usepackage{slashed}
\usepackage{color}
\usepackage{setspace}
\usepackage{hyperref}
\usepackage{empheq}

\newcommand{\ba}{\begin{array}}
\newcommand{\ea}{\end{array}}

\newcommand{\pert}{\mathrm{pert}}
\newcommand{\order}{\mathcal{O}}

\def\lQ{\Lambda_{\rm QCD}}
\newcommand{\nn}{\nonumber}
\newcommand{\be}{\begin{equation}}
\newcommand{\ee}{\end{equation}}
\newcommand{\bea}{\begin{eqnarray}}
\newcommand{\eea}{\end{eqnarray}}
\def\al{\alpha}

\def\siml{{\ \lower-1.2pt\vbox{\hbox{\rlap{$<$}\lower6pt\vbox{\hbox{$\sim$}}}}\ }} 
\def\simg{{\ \lower-1.2pt\vbox{\hbox{\rlap{$>$}\lower6pt\vbox{\hbox{$\sim$}}}}\ }} 

\newcommand{\MS}{\overline{\rm MS}}

\newcommand{\PV}{\rm PV}

\newcommand{\eq}[1]{Eq.~\eqref{#1}}

\begin{document}
\title{\vskip-3cm{\baselineskip14pt
}
\vskip1.5cm
The mass of the lightest gluelump}
\author{Cesar Ayala$^{1}$ and 
Antonio Pineda$^{2,3}$\\[0.5cm]
{\small ${}^1$  \it Departamento de Ingenier\'ia y Tecnolog\'ias, Sede La Tirana, Universidad de Tarapac\'a, 
}\\
{\small \it Av.~La Tirana 4802, Iquique, Chile} \vspace{0.3cm}\\
{\small ${}^2$ \it Institut de Física d'Altes Energies (IFAE),}
{\small\it The Barcelona Institute of Science and Technology,}\\
{\small\it Campus UAB, 08193 Bellaterra (Barcelona), Spain}\\
{\small ${}^3$ \it Grup de Física Teòrica, Dept. Física,}\\
{\small \it Universitat Autònoma de Barcelona, E-08193 Bellaterra, Barcelona, Spain}}
\date{}

\maketitle

\thispagestyle{empty}

\begin{abstract}
We consider QCD with $n_f=0$ or $n_f=3$ light quarks. 
We give the most up-to-date determinations of the normalization of the leading renormalons of the pole mass, the singlet static potential, the octet static potential, and the gluelump energy. They read $Z^{\MS}_m=-Z^{\MS}_{V_s}/2=\{0.604(17),0.551(20)\}$, $Z^{\MS}_{V_o}=\{0.136(8),0.121(13)\}$, and $Z^{\MS}_A=\{-1.343(36),-1.224(43)\}$, for $n_f=0$ and $n_f=3$ respectively. 
We obtain two independent renormalization group invariant and renormalization scale independent determinations of the energy of the ground state gluelump in the principal value summation scheme: $\Lambda_{B}^{\PV}=2.47(9)r_0^{-1}$  and $\Lambda_{B}^{\PV}=2.38(11)r_0^{-1}$ in the quenched approximation with $r_0^{-1} \approx 400$ MeV. They combine in 
$\Lambda_{B}^{\PV}=2.44(7)r_0^{-1}$.
\\[2mm]
\end{abstract}

\newpage
\index
\tableofcontents

\vfill
\newpage

\section{Introduction}
\label{int}

Heavy quarkonium hybrids and gluelumps are very interesting states for the study of the QCD dynamics. On the theoretical side, these systems open a venue to study the behavior of the QCD interactions in color representations different from the fundamental one (which is the one usually found in nature associated with quarks), with the significant simplification that the source of color behaves as, basically, a static source, unlike in studies of glueball states (the other natural possibility to study color representations different from the fundamental one). The proliferation of new exotic hadrons found in recent experiments \cite{Brambilla:2019esw} has only but increased the interest in them, since some of these exotic hadrons may be interpreted as heavy quarkonium hybrids.

A particular feature of heavy quarkonium hybrids and gluelumps is that they are related in the limit of small $r$, the distance between the heavy quark constituents of the heavy quarkonium hybrids. This is the situation we will consider in this paper. 

In Ref. \cite{Brambilla:1999xf}, the gluelumps and the short distance regime of the static hybrid energies were studied
within the static version of the effective field theory (EFT) potential NRQCD (pNRQCD)\cite{Pineda:1997bj,Brambilla:1999xf}. This study allowed to identify its general features. Some earlier results \cite{Hasenfratz:1980jv,Perantonis:1990dy,BBV,Foster:1998wu,Bali:2000gf} were recovered within a unified framework and in some cases extended. Using this EFT the static hybrid energies are organized in an expansion in powers of $r$.

One can go beyond this analysis and use lattice data plus the knowledge of the perturbative expansion of the
octet/singlet potential and of the static self-energy quark/gluelump masses (in the lattice scheme) to obtain numerical values for the gluelump masses and a more quantitative understanding of the hybrid static energies at short distances. Indeed, accurate lattice data of the heavy quarkonium hybrid static potentials \cite{Juge:1999ie,Bali:2003jq,Schlosser:2021wnr} and of the gluelump static energies \cite{Foster:1998wu,Marsh:2013xsa,Herr:2023xwg} is available. Moreover, the perturbative series expansions relevant to this paper have also been computed with increasing accuracy over the years. The precision reached nowadays is next-to-next-to-next-to-leading order (NNNLO) for fixed order computation and next-to-next-to-next-to-leading logarithmic (NNNLL) order for renormalization group (RG) improved computations in the $\MS$ for the static singlet potential/energy \cite{Fischler:1977yf,Schroder:1998vy,Brambilla:1999qa,Gorishnii:1991hw,Smirnov:2008pn,Anzai:2009tm,Smirnov:2009fh,Pineda:2000gza,Brambilla:2009bi,Pineda:2011db,Brambilla:2006wp,Pineda:2011aw,Lee:2016cgz}.\footnote{Fixed order computations trivially incorporate the resummation of soft logarithms by setting the renormalization scale to be proportional to $1/r$ up to a coefficient of order one. In this paper, the N$^k$LL RG improved expressions refer to those on which the resummation of ultrasoft logarithms is also implemented.} The static octet potential/energy has also been computed over the years. The difference with the singlet (up to an overall color factor) starts at two loops \cite{Kniehl:2004rk}. At three loops the expressions can be found in \cite{Anzai:2013tja,Lee:2016cgz}. The resummation of ultrasoft logarithms was obtained in \cite{Pineda:2000gza} with NNLL accuracy and in  \cite{Pineda:2011db} with NNNLL accuracy.

Another ingredient needed in our analyses is the perturbative expansion of the self energies of the fundamental and adjoint representations in the lattice scheme to high orders. They have been obtained in  \cite{Bauer:2011ws,Bali:2013pla,Bali:2013qla} to order 20 using Numerical Stochastic Perturbation Theory techniques  \cite{DRMMOLatt94,DRMMO94,DR0} and a careful infinite volume extrapolation limit.

This brief account clearly illustrates the impressive effort devoted to the computation of these quantities. Yet, even after this tremendous amount of perturbative computations and of lattice data accumulated over the years, the quantitative implementation of the operator product expansion formulas that follow from the application of EFTs to the factorization of scales one has in these systems is, however, not possible without extra theoretical input. The reason is that the perturbative series that appear in the analyses are asymptotic due to renormalons \cite{tHooft}. This is a general problem when different scales are factorized, and, in particular, when factorizing perturbative scales  from non-perturbative ones. In this last case, the bad convergence is just a manifestation of the fact that one cannot define non-perturbative quantities without defining their perturbative counterparts with exponential accuracy.

The lesson to be learned from the previous discussion is that, for quantitative studies, it is compulsory to treat (and define) the perturbative series with exponential accuracy. A first step to reach this goal is to quantitatively obtain the leading asymptotic behavior of the perturbative series that appear in the problem. This is tantamount to obtaining the closest singularities in the Borel plane of the Borel transform of the perturbative series. The location of the singularities and their strength are fixed completely by the Operator Product Expansion (OPE)/EFT analysis. On the other hand, the overall normalizations cannot be computed exactly. Nevertheless, they can be computed with increasing degree of accuracy the more terms in the perturbative expansion, and the better the structure of the OPE, are known. Indeed, there has been an ongoing effort over the years to compute them. Previous results for the normalizations that appear in this work can be found in Refs. \cite{Pineda:2001zq,Bali:2003jq,Bauer:2011ws,Bali:2013pla,Bali:2013qla,Ayala:2014yxa,Beneke:2016cbu}. We give more up to date results for these normalizations in Sec. \ref{Sec:Norm}.

The second step is to develop a scheme to handle the renormalon. The first such attempt in the context of gluelump/heavy quarkonium hybrids was made in Ref. \cite{Bali:2003jq}. A more updated value of the gluelump mass in the RS scheme can be found in Ref. \cite{Herr:2023xwg}. In these references, the threshold mass named RS (and RS') \cite{Pineda:2001zq} was used. This threshold mass explicitly subtracts the renormalon in the Borel plane, introducing a scale $\nu_f$ that acts as an infrared cutoff. Such infrared
cutoff kills the renormalon behavior of the perturbative series, producing a convergent perturbative
series but introducing a linear power-like dependence in an infrared cutoff $\nu_f$ of the gluelump masses and of the static single/octet potential.  Instead of following this path, we will use the hyperasymptotic expansion methods developed in \cite{Ayala:2019uaw,Ayala:2019hkn} since they do not introduce spurious power-like cutoff dependence, allowing us to obtain determinations of the nonperturbative constants that scale homogeneously in powers of $\Lambda_{\rm QCD}$ (i.e., they are of natural size and scale/scheme independent). Moreover, it is, in principle, possible to have a parametric control of the error of the truncation of the hyperasymptotic expansion. On top of that, we will use the more up-to-date perturbative expressions and lattice data available at present. 

\section{Gluelumps}
\label{Sec:gluelumps1}
Gluelump states are created by a static source in the octet (adjoint) representation attached
to some gluonic content (H) such that the state becomes a singlet under gauge
transformations. This is what would happen to heavy gluinos in the static approximation. One can then consider a heavy gluino effective theory. The bound state mass of a gluino and glue can then be written as an expansion in powers of $1/m_{\tilde g}$ in the following way (see, for instance, the discussion in Ref. \cite{Bali:2003jq}): 
\be
M_{H,\tilde G}=m_{\tilde g,\PV}+ \Lambda_{H}^{\PV} +{\cal O}(1/m_{\tilde g,\PV})
\,.
\ee
This is analogous to the $B$ meson mass in heavy quark effective theory (for a review see \cite{Neubert:1993mb}):
\be
\label{MB}
M_B=m_{\PV}+\bar \Lambda_{\PV} +{\cal O}(1/m_{\PV})
\,.
\ee
Evaluations in the lattice of the energy of a static adjoint source attached to glue yield the following analogous expressions: 
\be
\label{LambdaHL}
\Lambda_H^L(a)=\delta m_{A,\PV}^L(a)+\Lambda_{H}^{\PV}+{\cal O}(a^2)
\,.
\ee
Note that, in this last expression, we state that the lattice artifacts are ${\cal O}(a^2)$ instead of ${\cal O}(a)$. This is to be expected from symmetry arguments of the computation using the Wilson action, see the discussions in Refs. \cite{Foster:1998wu,Bali:2003jq,Herr:2023xwg}.\footnote{For $\bar \Lambda$ the situation is more uncertain, as there are always light quarks involved in the computation. Indeed, the analysis of Ref. \cite{Ayala:2019hkn} shows sizeable lattice artifacts after subtracting perturbation theory, though the nature of these is not clear. In the present paper, lattice artifacts are much smaller, and difficult to disentangle from the uncertainties of the hyperasymptotic expansion of the perturbative term.}
 
The different terms in the above expressions in the $1/m$ expansion (or $a$ expansion) are ill defined. This is the reason we use the Principal Value (PV) summation scheme for the inverse Borel transform of the perturbative series to define the (otherwise divergent) perturbative sum of the first term of the $1/m$ expansion (or $a$ expansion). In turn, this defines the leading nonperturbative constants: 
$\bar \Lambda_{\PV}$ or $ \Lambda_{H}^{\PV}$ that appear in these expansions. What it is important to stress is that, once these constants appear, they are universal. They do not depend on how they were obtained, neither on the scale or scheme used  for $\alpha$. 

For the purpose of this paper, we focus on Eq. (\ref{LambdaHL}), since we will determine $ \Lambda_{H}^{\PV}$ by fitting our theoretical expressions to lattice data. 

The perturbative expression for $\delta m_{A,\PV}^L(a)$ is not known exactly, nor, consequently, is known the PV summation of it. We evaluate it approximately using the hyperasymptotic expansion, which reads
\be
\label{deltamAPV}
\delta m_{A,\PV}^L(a)=\delta m_{A}^{(P)}(1/a)+\frac{1}{a}\Omega_A(1/a;a)+\sum_{N_P+1}^{N'=3 N_P}\frac{1}{a}[c_{A,n}-c_{A,n}^{\rm (as)}]\alpha_L^{n+1}(a)+{\cal O}(a^2)
\ee
for our present level of precision. 

Crucial for this analysis is the determination of the coefficients of the perturbative expansion in the (Wilson action) lattice scheme strong coupling, $\alpha_L$, obtained in Refs. \cite{Bauer:2011ws,Bali:2013pla,Bali:2013qla}. 

We now turn to explicitly write the different terms that appear in Eq. (\ref{deltamAPV}). The leading term reads
\be
\label{deltamASuper}
\delta m_{A}^{(P)}(1/a)=\frac{1}{a}\sum_{n=0}^{N_P}c_{A,n}\alpha_L^{n+1}(a)
\,,
\ee
where
\be
N_P=\frac{2\pi}{\beta_0\alpha_L(a)}\left(1-c\alpha_L(a)\right)
\,.
\ee
and $c$ is a number of order one to ensure that $N_P$ is integer. By default, we will be take the smallest possible positive value of $c$ that fulfils that $N_P$ is an integer. 

Note that, by working in the lattice scheme, $\alpha_L(a)$ is much smaller than the corresponding one in the $\MS$ scheme. In turn, this means that one has to go to higher orders in the perturbative expansion to reach the asymptotic regime of the perturbative series (in our case, $N_P$ is around 6,7). 

Keeping only Eq. (\ref{deltamASuper}) corresponds to having superasymptotic \cite{BerryandHowls}
precision, which is obtained by truncating the perturbative expansion at the minimal term. To improve on that, we have to add the leading terminant \cite{Dingle}. Its expression reads (see Refs. \cite{Ayala:2019uaw,Ayala:2019hkn} for extra details)
\be
\label{OmegaA}
\Omega_A(\nu; a)=Z^X_{\delta m_A} \Omega_X(\nu; a)
\,,
\ee
where\footnote{Note that, unlike in Refs. \cite{Ayala:2019uaw,Ayala:2019hkn}, we do not include the normalization of the renormalon in the constant $K^{(P)}$ (see Eq. (\ref{KP})). The reason is that $\Omega_X$ is the same for several quantities: the pole mass, the static singlet/octet potential, and the gluelump energy.}
\be
\label{eq:OmegaV}
\Omega_X(\nu; a)
=\sum_{j=0}^{\infty}c_j\Delta \Omega (b-j)
\,,
\ee
and $b=\beta_1/(2\beta_0^2)$. 
The coefficients $c_k$ are pure functions of the $\beta$-function coefficients, as first shown in \cite{Beneke:1994rs} for the case of the pole mass. They can be found in \cite{Beneke:1998ui,Pineda:2001zq,Ayala:2014yxa}. At low orders they read ($c_0=1$) 
\be
\label{eq:c1c2c3}
c_1=s_1 \,,\quad 
c_2=\frac{1}{2}\frac{b}{b-1}(s_1^2-2s_2)
\,,
\quad
c_3=\frac{1}{6}\frac{b^2}{(b-2)(b-1)}(s_1^3-6s_1s_2+6s_3)
\,,
\ee
 where the $s_n$ coefficients are defined in \cite{Ayala:2019uaw}. Note that the beta coefficients correspond to those in the renormalization scheme used for $\alpha$, but the expression is valid in any scheme.
Finally,  
\be
\Delta \Omega (b-j)=\nu a\frac{1}{\Gamma(1+b)}\left(\frac{\beta_0}{2\pi}\right)^{N_P+1}\alpha_X^{N_P+2}(\nu)
\int_{0,\PV}^{\infty} dx \frac{x^{b+N_P+1}e^{-x}}{1-x\frac{\beta_0 \alpha_X^{N_P+2}(\nu)}{2\pi}}
\,.
\ee
This expression is amenable to a numerical analysis. It also performs a partial resummation of powers of $\alpha_X$. If one performs a strict weak coupling analysis, Eq. (\ref{eq:OmegaV}) can be approximated to 
\be
\label{eq:OmegaV2}
\Omega_X(\nu; a)=\sqrt{\alpha_X(\nu)}K^{(P)}
\nu a \, 
e^{-\frac{2\pi}{\beta_0 \alpha_X(\nu)}}
\left(\frac{\beta_0\alpha_X(\nu)}{4\pi}\right)^{-b}
\bigg(1+\bar K_{1}^{(P)}\alpha_X(\nu)+\bar K_{2}^{(P)}\alpha_X^2(\nu)+\mathcal{O}\left(\alpha_X^3(\nu)\right)\bigg)
\,,
\ee
\bea
\label{KP}
K^{(P)}&=&-\frac{ 2^{1-b}\pi}{\Gamma(1+b)}\beta_0^{-1/2}\bigg[-\eta_c+\frac{1}{3}\bigg]
\,,
\\
\bar K_{1}^{(P)}&=&\frac{\beta_0/(\pi)}{-\eta_c+\frac{1}{3}}\bigg[-b_1 b \left(\frac{1}{2}\eta_c+\frac{1}{3}\right)
-\frac{1}{12}\eta_c^3+\frac{1}{24}\eta_c-\frac{1}{1080}\bigg]
\,,
\\
\bar K_{2}^{(P)}&=&\frac{\beta_0^2/\pi^2}{-\eta_c+\frac{1}{3}}
\bigg[-w_2 (b -1) b \left(\frac{1}{4}\eta_c+\frac{5}{12}\right)
+b_1b\left(-\frac{1}{24}\eta_c^3-\frac{1}{8}\eta_c^2
-\frac{5}{48}\eta_c-\frac{23}{1080}\right)
\nn
\\
&&
-\frac{1}{160}\eta_c^5
-\frac{1}{96}\eta_c^4+\frac{1}{144}\eta_c^3
+\frac{1}{96}\eta_c^2-\frac{1}{640}\eta_c-\frac{25}{24192}\bigg]
\,,
\eea
and so on (see \cite{Ayala:2019uaw}).  In particular,
\be
\eta_c=-b +\frac{2\pi   c}{\beta_0}-1 \;, \quad b_1= s_1 \;, \quad {\rm and} \quad w_2=\left(\frac{s_1^2}{2}-s_2\right)\frac{b}{b-1}
\,.
\ee
 In this section we take $X=L$.

Note that the left-hand side of Eq. (\ref{OmegaA}) is basically independent of scheme but not completely. 

To go beyond the leading terminant, we have to add the term 
\be
\label{DeltaM}
\displaystyle{\sum_{N_P+1}^{N'=3N_P}\frac{1}{a}[c_{A,n}(1/a)-c_{A,n}^{\rm (as)}(1/a)]\alpha_L^{n+1}(a)}
\,,
\ee
where the leading asymptotic behavior of the perturbative coefficients reads
\be
\label{rnas}
c_{A,n}^{\rm (as)}(\nu)=Z^X_{\delta m_A} \nu a \,\left({\beta_0 \over
2\pi}\right
)^n \,\sum_{k=0}^\infty c_k{\Gamma(n+1+b-k) \over
\Gamma(1+b-k)}
\,.
\ee
The coefficients $c_k$ have been defined in Eq. (\ref{eq:c1c2c3}). Again, this expression holds in any renormalization scheme. The rate of convergence in the $1/n$ expansion could be different, though, depending on the scheme. Indeed, one finds a faster convergence in the $\MS$ scheme compared with the lattice scheme. 

\begin{figure}
\begin{center}
\includegraphics[width=0.814\textwidth]{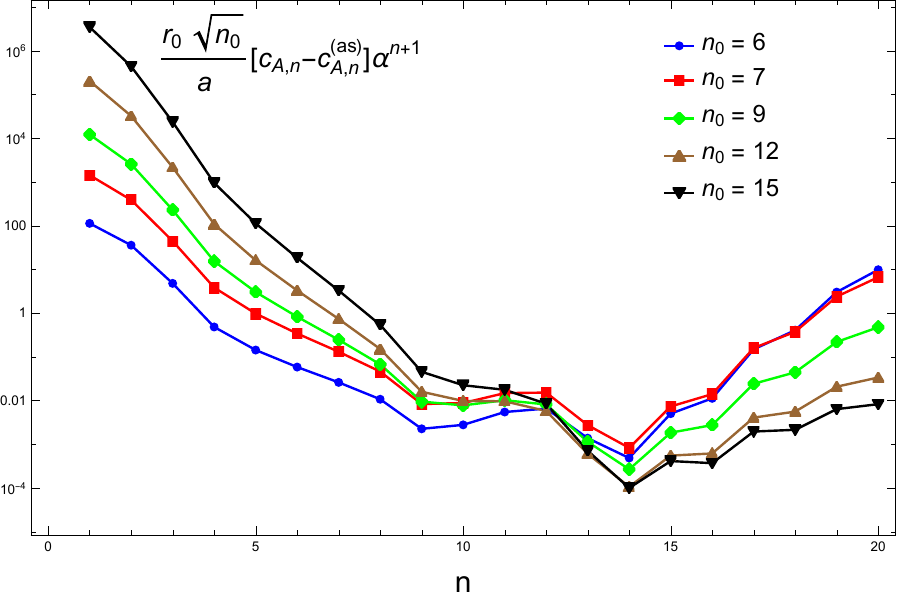}
\end{center}
\caption{Plot of $\sqrt{n_0}\frac{1}{a}[c_{A,n}(1/a)-c_{A,n}^{\rm (as)}(1/a)]\alpha_L^{n+1}(a)$ in $r_0$ units for different values of $n_0$. See the main text for details.}
\label{Fig:TestOrderLatt}
\end{figure}

To reliable compute Eq. (\ref{DeltaM}), a very fine tune cancellation between the exact and the asymptotic coefficients is needed. To illustrate the problem, we plot in Fig. \ref{Fig:TestOrderLatt} the following quantity:
\be
\sqrt{n_0}\frac{1}{a}[c_{A,n}(1/a)-c_{A,n}^{\rm (as)}(1/a)]\alpha_L^{n+1}(a)
\,,
\ee
where $\alpha_L^{n+1}(a)$ is fixed in terms of $n_0$ using the following equality
\be
n_0=\frac{2\pi}{\beta_0\alpha_L(a)}
\,.
\ee
Given $\alpha_L^{n+1}(a)$, the expression of $a$ in term of $r_0$ is fixed using the Necco-Sommer formula \cite{Necco:2001gh} for $n_0=6,7$, and the perturbative expression (see, for instance, Eq. (4) in Ref. \cite{Ayala:2019uaw}) for $n_0=9,12,15$. This figure is in close analogy with Fig. 15 in Ref. \cite{Bali:2013pla}, and we can draw the following conclusions: If there were a renormalon at $N' \sim 2n_0$, one would expect the minimal term to be located at around $2n_0$ and, accordingly, to change for different values of $\alpha_L$. Instead, what we observe is that the minimum is always located at the same $n_0$. Another smoking gun prediction of a renormalon at $N' \sim 2n_0$ is that the minimal term should scale as $\Lambda_{\rm QCD}^3 a^2$. Again, we do not see such behavior. As a final remark, if we move away from the value of $Z_A$ deduced in Eq. (\ref{Zanf0bis}), the minimal term gets closer to $n_0$, indicating that there is a leftover contribution proportional to the leading renormalon in Eq. (\ref{DeltaM}) (albeit multiplied by a small number). 

Our conclusion from this analysis is that there is no renormalon at $N' \sim 2N_P$, and that what we observe is a limitation in the precision of the coefficients. This implies that, with the exponential precision required in the computation, adding terms beyond $N' \sim 13$ would deteriorate the accuracy of the theoretical prediction. Therefore, this will be the maximal number of terms that we will introduce in our evaluations. In practice, this number is not much different from $\sim 2N_P$ but we take this as a numerical accident. 

The previous discussion, and theoretical arguments, rule out ${\cal O}(a)$ effects. This is the reason Eqs. (\ref{LambdaHL}) and  (\ref{DeltaM}) assume that the unknown corrections are of order $a^2$ (if these are related with renormalons, this would mean that $N'$ should be of order $3N_P$).  On top of that, there could be nontrivial anomalous dimensions. Overall, the leading unknown term is expected to scale as $\Lambda^3 a^2$ times a, comparatively smooth, function in powers of $\alpha$. 

\section{Static hybrids}
The energy of a static quark and a static antiquark in a colour singlet configuration separated by a distance $r$, $E_s(r)$, admits an OPE using pNRQCD \cite{Pineda:1997bj,Brambilla:1999xf}: 
\be
\label{Es1}
E_s(r)=2m_{\PV}+V_{s}^{\PV}(r;\nu_{us})+ \delta E^{\PV}_{s,us}(r;\nu_{us})
\,.
\ee
The hyperasymptotic expression for $m_{\PV}$ can be found in Ref. \cite{Ayala:2019hkn}, and it will not be discussed further in this paper. $ V_s(r;\nu_{us})$ only encodes the physics associated with the scale $1/r$, and the resummation of large logarithms associated with the ultrasoft scale.  Its hyperasymptotic expansion has been studied in detail in Ref. \cite{Ayala:2020odx}. $\delta E_{s,us}(r;\nu_{us})$ encodes the physics associated with scales smaller than $1/r$ and appears at ${\cal O}(r^2)$ in the multipole expansion. Its hyperasymptotic expansion has also been studied in Ref. \cite{Ayala:2020odx}.

The energy of a static quark and a static antiquark in a colour octet configuration separated by a distance $r$, $E_H(r)$, follows a similar pattern, but there are significant differences. It can still be splitted into the effects associated with the soft scale $\nu_s \sim 1/r$ and the effects associated with smaller scales in a formula analogous to the previous one:
\be
\label{EH1}
E_H(r)=2m_{\PV}+V_{o}^{\PV}(r;\nu_{us})+\delta E^{\PV}_{o,us}(r;\nu_{us})
\,.
\ee
 The physical effects associated with the scale $1/r$ are encoded in the octet potential $V_o$, where, again, the resummation of large logarithms associated with the ultrasoft scale is also incorporated. The $r$ distances considered in this paper are small compared with $1/\Lambda_{\rm QCD}$. Therefore, an hyperasymptotic evaluation of the octet potential, $V_o$, is possible. The most up-to-date pure perturbative analysis yields expressions for this quantity with NNNLL precision in the $\MS$ scheme. The perturbative expression of this quantity is explained in detail in Ref. \cite{Pineda:2011db}.
The asymptotic expansion of $V_{o}^{\rm PV}$ coincides, by construction, with the perturbative expression of $V_o$:
\be
V_{o}^{\rm PV}(r;\nu_s;\nu_{us}) \sim \sum_{n=0}^{\infty} v^{(o)}_n \al_{\MS}^{n+1}(\nu_s)
\,.
\ee
The natural scale for this expansion is $\nu_s \sim 1/r$. Similarly to the singlet potential, the octet potential is logarithmic infrared divergent. Such behavior first appears in $v_3^{(o)}$, which endures a linear $\ln (\nu_{us})$ dependence that is not absorbed in the renormalization of the strong coupling constant in $V_o^{\rm PV}$. This logarithmic behavior is cancelled instead by $\delta E_{us}$. Therefore, both $V_{\rm PV}(\nu_{us})$ and $\delta E_{us}(\nu_{us})$  are renormalization scale and scheme dependent in perturbation theory. Such scale and scheme dependence cancels in the sum in \eq{EH1}. These effects call for the resummation of large 
powers of $\al(\nu_s)\ln (\frac{\nu_s}{\nu_{us}})$ (we take $\nu_s \sim 1/r$). These can be resummed using RG techniques. We incorporate these effects in $\delta V_o^{\rm RG}$. Note that this effect is ${\cal O}(r^2)$ suppressed in the multipole expansion. In a weak coupling analysis, this reflects in an ${\cal O}(\al^2)$ suppression plus possible extra powers of $\alpha$ associated with the ultrasoft scale.

Overall, the RG improved potential reads
\be
V_o^{\rm PV}=V_{\rm PV}(r;\nu_s;\nu_{us}=\nu_s)+\delta V_o^{\rm RG}(r;\nu_s,\nu_{us})
\,.
\ee
The expression for the perturbative expression of $V_o^{\rm PV}(r;\nu_{us}=\nu_s)$ with NNNLO accuracy can be found in Eqs. (39) and (40) of Ref. \cite{Pineda:2011db}. The value of $a_3^{(o)}$ is not available in that reference. It corresponds to $a_3^{[8]}$ in Ref. \cite{Anzai:2013tja} (see their Eq. (21)). 

The expression for $\delta V_o^{\rm RG}(r;\nu_s,\nu_{us})$ can be found in Eq. (30) of Ref. \cite{Pineda:2011db} with NNNLL accuracy. This object is expected to be renormalon free.

\subsection{Hyperasymptotic approximation of $V_o$}

We now focus on the hyperasymptotic expansion of $V_o^{\rm PV}(r;\nu_s;\nu_{us}=\nu_s)$. We construct the hyperasymptotic expansion of $V_o^{\rm PV}$ along the lines of Refs. \cite{Ayala:2019uaw,Ayala:2019hkn,Ayala:2020odx}. 
$V_o$ does not have ultraviolet renormalons, whereas the leading infrared ones are located at dimensions 1 and 3 (i.e. at $u=1/2$ and at $u=3/2$ in the Borel plane). The termination of the perturbative series associated with these renormalons produces nonperturbative power contributions of order $\lQ$ and $r^2 \lQ^3$ respectively:
\be
\label{eq:VPV}
V_o^{\rm PV}=V_{o,P}+\frac{1}{r}\Omega_{V_o}+\sum_{n=N_P+1}^{3N_P/N_{max}} (v^{(o)}_n-v_n^{(o,\rm as)}) \al_{\MS}^{n+1}(\nu_s)
+O(\lQ^3 r^2)
\,,
\ee
where $N_{max}=3$, (in the coefficients $ v^{(o)}_n$ we set $ \nu_{us}=\nu_s$) 
\be
V_{o,\rm P}\equiv \sum_{n=0}^{N_P} v^{(o)}_n \al_{\MS}^{n+1}(\nu_s)\; ,
\ee 
and
\be
\label{eq:NP}
N_P=\frac{2\pi}{\beta_0\al_{\MS}(\nu_s)}\left(1-c\al_{\MS}(\nu_s)\right)
\,.
\ee
Approximating $V_o^{\rm PV}$ by $V_{o,\rm P}$ corresponds to achieving superasymptotic precision. In the generic labeling $(D,N)$ of the truncations of the hyperasymptotic approximation defined in \cite{Ayala:2019uaw,Ayala:2019lak}, it corresponds to $(0,N_P)$ precision. The next order in the hyperasymptotic approximation is labeled as (1,0) and means adding $\frac{1}{r}\Omega_{V_o}$ to $V_{o,\rm P}$. Its explicit expression reads 
\begin{equation}
\label{eq:OmegaVo}
\Omega_{V_o}=Z_{V_o}\Omega
\,,
\end{equation}
where $\Omega$ has been defined in Eq. (\ref{eq:OmegaV}) (but note that now we work in the $\MS$ scheme).

We now move beyond (1,0) hyperasymptotic precision by including the third term in \eq{eq:VPV}, where 
the leading asymptotic behavior of $v_n^{(o)}$ reads
\be
\label{vnas}
v_n^{(o,\rm as)}(\nu_s)=Z_{V_o} \nu_s \,\left({\beta_0 \over
2\pi}\right
)^n \,\sum_{k=0}^\infty c_k{\Gamma(n+1+b-k) \over
\Gamma(1+b-k)}
\,,
\ee
where $b=\beta_1/(2\beta_0^2)$ and the coefficients $c_k$ are the same as in Eq. (\ref{eq:c1c2c3}). 
By including the third term in \eq{eq:VPV}, one then reaches hyperasymptotic precision (1,$N_{max}/3N_P$). 

If we have enough terms in perturbation theory, we may start to be sensitive to the ${\cal O}(r^2)$ ($u=3/2$) renormalon. This will happen if $N_{max} \sim 3N_P$. This is the maximal accuracy we will explore in this paper. 

In the singlet static  potential case, the ${\cal O}(r^2)$ renormalon-associated ambiguity is completely correlated with the ${\cal O}(r^2)$ ultrasoft contribution proportional to $V_A^2$, where $V_A$ is the Wilson coefficient of the leading singlet-octet mixing term that appears in the pNRQCD Lagrangian.\footnote{Not to be confused with the adjoint static potential defined later.}
 $V_A$ is known to be 1 with NLL accuracy \cite{Pineda:2000gza,Brambilla:2009bi}. For the hybrid static energies, there are two other possible terms that contribute to ${\cal O}(r^2)$. One is proportional to $V_B^2$ and the other to $V_C$ (the operators of the effective theory can be found in Eqs. (1) and (2) of Ref. \cite{Pineda:2011db}). Within perturbation theory, these terms are zero because there is no scale to regulate them (unlike the term proportional to $V_A^2$, which scales with the ultrasoft scale $\Delta V\equiv V_o-V_s$). Nevertheless, they may produce effects on the asymptotic behavior of the perturbative expansion of the octet potential. These two extra terms can have an infrared and ultraviolet renormalon. The ultraviolet ones would cancel with the corresponding infrared ones of the octet potential.  Therefore, they will affect the asymptotic behavior of the ${\cal O}(r^2)$-associated renormalon. On top of that, the values of $V_B$ and $V_C$ are only known to leading order in $\alpha$, i.e.,  $V_B=V_C=1+{\cal O}(\alpha)$. Therefore, we refrain from trying to obtain the normalization of these subleading renormalons, nor incorporating these effects in our determinations of the gluelump masses. The effect of these renormalons would be to add a terminant proportional to $r^2$ (times a smooth function in $\alpha$), which would mix with genuine nonperturbative/ultrasoft effects we discuss in the next subsection. The most we will do is to explore whether the perturbative expansion of the octet potential is sensitive to these renormalons. This is postponed to Sec.  \ref{Sec:ZFmain}. 

\subsection{Nonperturbative and ultrasoft scales}

Whereas the soft scale and the associated logarithmic running can confidently be treated using weak coupling analyses, the situation at scales of order $2/r_0$ ($ \sim 800$ MeV) is more uncertain. Still, physics at those scales can be organized in powers of $r$ by using the multipole expansion, as implemented in the static version of the pNRQCD effective theory. This is true even in the event that we cannot make a perturbative evaluation of those corrections. Unlike in the singlet case, there is a contribution at ${\cal O}(r^0)$. This leading order term is $r$ independent and will actually correspond to the gluelump mass: $\Lambda_{H}^{\PV}$. Therefore, it is a purely nonperturbative (it will vanish to all orders in perturbation theory) quantity that scales linearly in $\Lambda_{\rm QCD}$. The next correction in the multipole expansion is of ${\cal O}(r^2)$: $\delta E^{(2)\PV}_{o,us}(r;\nu_{us})$. This last quantity was computed in perturbation theory in \cite{Pineda:2011db} with NNNLO accuracy (see Eq. (42) in that reference). Note such expression has some imaginary terms, reflecting the fact that static hybrids can decay in perturbation theory. We will set these purely imaginary terms to zero in this paper. If one has to consider nonperturbative effects in $\delta E^{(2)\PV}_{o,us}(r;\nu_{us})$, the situation is more uncertain (and proportional to three different Wilson coefficients: $V_A$, $V_B$ and $V_c$). As this term is at the limit of the precision we will be able to achieve, we will only use it for error analysis, either by taking its leading nonvanishing perturbative expression for the estimate:
\be
\label{dEUSpert}
\delta E^{(2)\PV}_{o,us}(r;\nu_{us})
\simeq
\frac{1}{2N_c}r^2(-\Delta V)^3\left(-\frac{\al(\nu_{us})}{9\pi}\right)
\left(
6\ln\left[\frac{\Delta V}{\nu_{us}}\right]+6\ln 2-5
\right)
\,,
\ee
or, alternatively, using
\be
\label{dEUSnopert}
\delta E^{(2)\PV}_{o,us}(r;\nu_{us})
\simeq
Ar^2
\,,
\ee
if we, instead, assume that the ultrasoft scales are in the nonperturbative regime.

Overall, we recast the formula we use for the spectrum of the hybrid static energy in the following way:
\be
\label{EH2}
E_H(r)=2m_{\PV}+V_{o}^{\PV}(r;\nu_{us})+\Lambda_{H}^{\PV}+\delta E^{(2)\PV}_{o,us}(r;\nu_{us})
\,.
\ee

We will confront the above theoretical expression with nonperturbative evaluations obtained from lattice simulations. 
If we consider lattice analyses, the following ``observables'' show up:
\be
E^L_{\Sigma_g^+}(r;a)=V_s^L(r;a)+{\cal O}(r^2)
\,,
\ee
\be
E^L_H(r;a)= V_o^L(r;a)+\Lambda^L_H+{\cal O}(r^2)
\,.
\ee
It is possible to relate these expressions with Eqs. (\ref{Es1}) and (\ref{EH1}) by following a discussion analogous to the discussion in Ref. \cite{Bali:2003jq} changing RS by PV (two different ways to treat the renormalon). In this paper, rather than considering each static energy independently, we consider the following combination:
\be
\label{EPiuSigmag}
E_{\Pi_u}-E_{\Sigma_g^+}=V_{A}^{\PV}+\Lambda_{H}^{\PV}+\delta E^{(2)\PV}_{A,us}
\,.
\ee
In the left hand side, we have dropped the index $L$ because it is possible to get the continuum limit of this energy difference. More importantly, in the right hand side, we can fully work in the $\MS$ scheme, where we have used the following definitions: $V_{A}^{\PV}=V_{o}^{\PV}-V_{s}^{\PV}$ and $\delta E^{(2)\PV}_{A,us}=\delta E^{(2)\PV}_{o,us}-\delta E^{\PV}_{s,us}$.

\section{Normalization of renormalons}
\label{Sec:Norm}

The theoretical expressions  for the observables written in the previous sections produce relations between the normalization of the leading renormalons associated with the different terms that appear in the OPE of them. From Eqs. (\ref{Es1}), (\ref{EH1}), (\ref{MB}) and (\ref{LambdaHL}), we obtain (see, for instance, Refs. \cite{Pineda:2001zq,Bali:2003jq})
\be
\label{ZmZVs}
2Z_m+Z_{V_s}=0
\,,
\ee
\be
\label{ZmZVoZA}
2Z_m+Z_{V_o}+Z_A=0\,,
\ee
\be
\label{ZmZbarL}
Z_m+Z_{\bar \Lambda}=0 
\,,
\ee
and
\be
\label{ZdeltamA}
Z_{\delta m_A}+Z_A=0
\,.
\ee
On top of that, we also have the relations $Z_{\delta m_A}=Z_{m_{\tilde g}}$ and $Z_{\delta m}=Z_{m}$.

Besides these relations, absolute values for the normalizations can be obtained if enough orders in the perturbative expansion of the associated term in the OPE are known. The above relations also give possible alternative ways to determine these normalizations. We will proceed to compute them in the following subsections. We will compute them by considering the ratio of the exact coefficient and the associated asymptotic expression, similar to how it was done in Refs. \cite{Bali:2013pla,Ayala:2014yxa}, to which we refer for details. In Ref. \cite{Bali:2013pla}, it was shown that this method yields better results for $Z_A$ and $Z_m$ than older methods, such as those used in Ref. \cite{Pineda:2001zq}.

\subsection{Revisiting $Z_{V_s}$ and $Z_m$}

Values of $Z_{V_s}$ and $Z_m$ have already been obtained in the past \cite{Ayala:2014yxa,Beneke:2016cbu}. It is the aim of this subsection to summarize and improve (when possible) these determinations in order to give the most up-to-date, and accurate, determination of $Z_{V_s}$ and $Z_m$. We will treat them as a single quantity since they are related by Eq. (\ref{ZmZVs}).

There are three quantities the perturbative expansions of which are known with enough accuracy to be sensitive to the associated $u=1/2$ renormalon. The three of them share the same infrared behavior but different ultraviolet and subleading renormalons. Therefore, we can consider them as independent determinations of the normalization and combine them to get a more accurate number. These three quantities are the static singlet potential, the pole mass, and the energy of a static quark in the fundamental representation. We now discuss them in turn. 

\begin{figure}[htb] 
\begin{minipage}[b]{.49\linewidth}
  \centering\includegraphics[width=84mm]{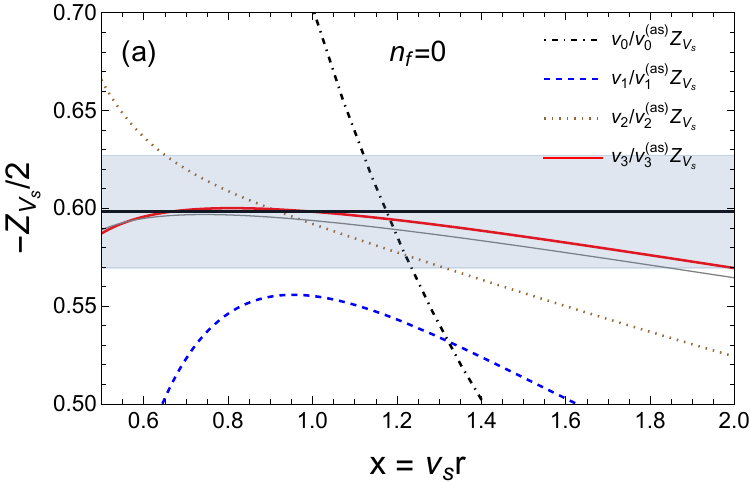}
  \end{minipage}
\begin{minipage}[b]{.49\linewidth}
  \centering\includegraphics[width=78mm]{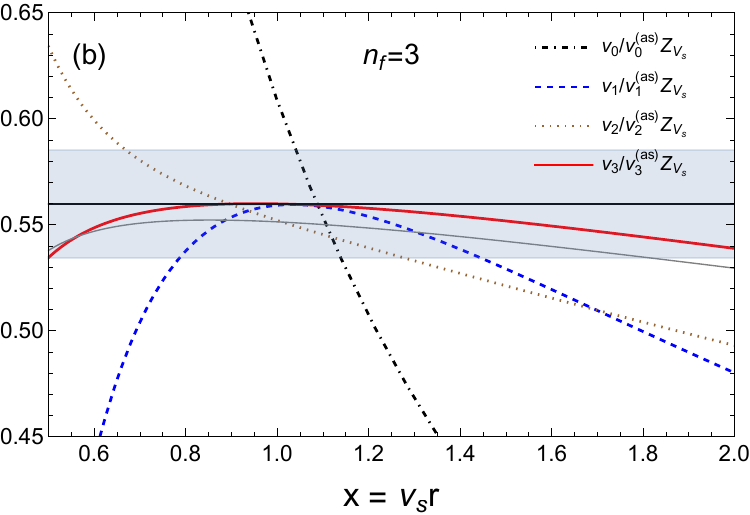}
\end{minipage}
\caption{\it Determination of $-Z_{V_s}/2$ using $v_n/v_n^{(as)}Z_{V_s}$ as a function of $x=\nu_s r$ and for different values of $n$ in the $\MS$ scheme. The gray continuous line is  $v_3/v_3^{(as)}Z_{V_s}$ withouth the ultrasoft logarithmically related term (see Ref. \cite{Ayala:2014yxa} for details). The black horizontal line is our final prediction and the blue band our final error estimate. {\bf (a)} are determinations with $n_f=0$ and {\bf (b)} with $n_f=3$.} 
\label{fig:ZVs}
\end{figure}

The most up-to-date determination of the normalization from the static potential was made in Ref. \cite{Ayala:2014yxa}. We have little to add here. The only difference is that the value of $\beta_4$ (the five-loop beta coefficient of the $\beta(\alpha)$ function) is now known \cite{Baikov:2016tgj}. This changes very little the result. Nevertheless, we repeat the analysis using the same error analysis as in that reference for completeness. We obtain
\bea
\label{ZVsnf0}
\left.-\frac{Z_{V_s}}{2}\right|_{n_f=0}&=&0.598(29)
\,,
\\
\label{ZVsnf3}
\left.-\frac{Z_{V_s}}{2}\right|_{n_f=3}&=&0.560(25)
\,,
\eea
and in Fig. \ref{fig:ZVs} we draw the analogous to Figs. 4 and 5 in Ref. \cite{Ayala:2014yxa}. 

Some considerations regarding the errors (a detailed account of the error analysis can be found in Ref. \cite{Ayala:2014yxa}):
\begin{itemize}
\item The ${\cal O}(1/n^3)$ term in the asymptotic coefficient is now known exactly since now the coefficient $\beta_4$ is known. This slightly modified the $x \equiv \nu_s r$ dependence.
\item $\nu_{us}=\nu_s$ in the singlet potential.
\item The central value is chosen at $x=1$. The final error is the largest number among the difference between $v_3/v_3^{(as)}Z_{V_s}$ and $v_2/v_2^{(as)}Z_{V_s}$  at $x=1$, and the one associated to scale variation. We estimate the latter by taking the difference with the values obtained if one takes $x=1/2$ or $x=2$. This error is then combined in quadrature with the error of the order $1/n^3$ correction, which is indeed very small.
\end{itemize}

\begin{figure}[htb] 
\begin{minipage}[b]{.49\linewidth}
  \centering\includegraphics[width=84mm]{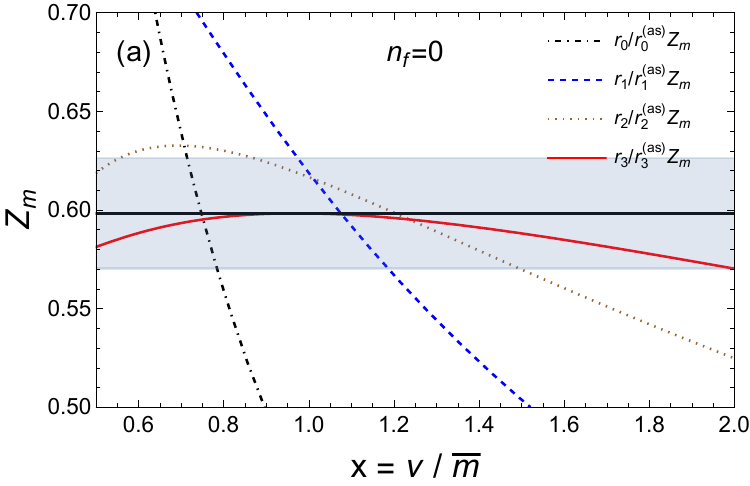}
  \end{minipage}
\begin{minipage}[b]{.49\linewidth}
  \centering\includegraphics[width=78mm]{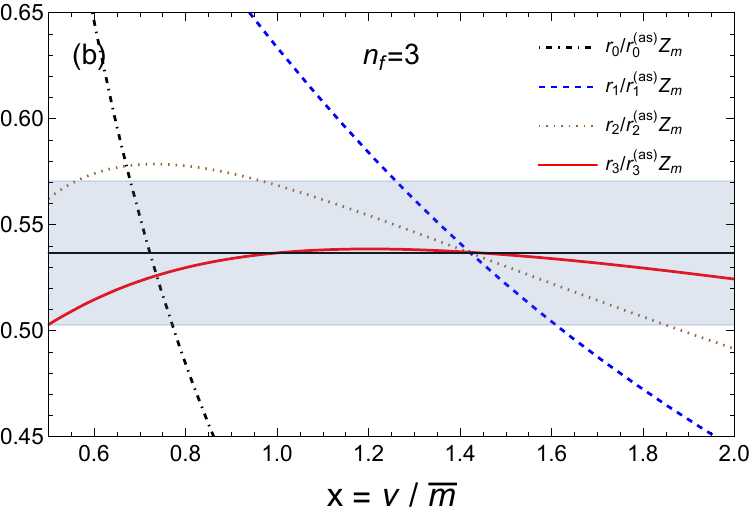}
\end{minipage}
\caption{\it Determination of $Z_{m}$ using $r_n/r_n^{(as)}Z_{m}$ as a function of $x=\nu/{\overline m}$ and for different values of $n$ in the $\MS$ scheme. The black horizontal line is our final prediction and the blue band our final error estimate. {\bf (a)} are determinations with $n_f=0$ and {\bf (b)} with $n_f=3$. For extra details see the main text and Ref. \cite{Ayala:2014yxa}.} 
\label{fig:Zm}
\end{figure}

We now turn to the mass. Compared with the analysis made in Ref. \cite{Ayala:2014yxa}, an extra term in the perturbative expansion of the pole mas
 is now known \cite{Marquard:2015qpa}. We incorporate this term and perform the error analysis in the same way as we did before for the case of the static singlet potential (except for the fact that there are no ultrasoft corrections). We obtain
\bea
\label{Zmnf0}
\left.Z_m\right|_{n_f=0}&=&0.598(28)
\,,
\\
\label{Zmnf3}
\left.Z_m\right|_{n_f=3}&=&0.537(34)
\,,
\eea
and in Fig. \ref{fig:Zm} we draw the analogous to Fig. 6 in Ref. \cite{Ayala:2014yxa}. 

Finally, there is also a determination from the computation of the energy of a static source in the fundamental representation to large orders in perturbation theory in the lattice (Wilson action) scheme 
\cite{Bauer:2011ws,Bali:2013pla,Bali:2013qla}:
\be
Z_m(n_f=0)=0.620(35)
\,.
\ee 

These three (two) determinations for $n_f=0$ ($n_f=3$) have different systematic errors.  They come from different perturbative expressions that only share in common that their leading renormalon is the same. Therefore, it makes sense to combine them in quadrature. If we do so, we obtain
\be
Z_m(n_f=0)=0.604(17)
\,,
\ee
\be
Z_m(n_f=3)=0.551(20)
\,.\ee

As a final remark, it is worth mentioning that, for the static potential, the subleading renormalon is located at $u=3/2$ in the Borel plane, whereas for the pole mass one expects them to be located at $u=1$ (infrared) and $u=-1$ (ultraviolet). So, in principle, the static potential is, theoretically, the best choice to determine the normalization of the pole mass. In practice, though, one gets similar results with similar errors. This may also indicate that there is no $u=1$ infrared renormalon in the pole mass (see the discussion in Ref. \cite{Ayala:2019hkn}), and that the $u=-1$ ultraviolet renormalon is not numerically important (at least to the level of precision one has at present). For the determination coming from the energy of a static source in the fundamental representation, the situation is not that clear: $|u|=1$ renormalons may exist for finite light quark masses. For zero light quark mass, one may expect these possible effects to be small.\footnote{A. Pineda acknowledges discussion on this point with Gunnar Bali.}

\subsection{Determination of $Z_{V_o}$ and $Z_A$}

The first determination of $Z_{V_o}$ was obtained in \cite{Bali:2003jq}. The perturbative expansion is now known with higher accuracy. As far as the determination of the normalization of the leading renormalon is concerned, the soft part of the perturbative expansion is known to one more order. On top of that, we also change the method for the determination of the normalization, and follow the same method we have used in the previous section.  

\begin{figure}[htb] 
\begin{minipage}[b]{.49\linewidth}
  \centering\includegraphics[width=84mm]{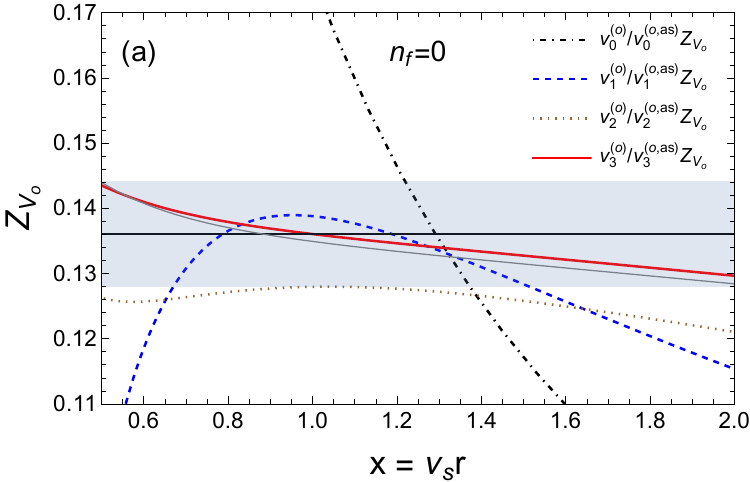}
  \end{minipage}
\begin{minipage}[b]{.49\linewidth}
  \centering\includegraphics[width=78mm]{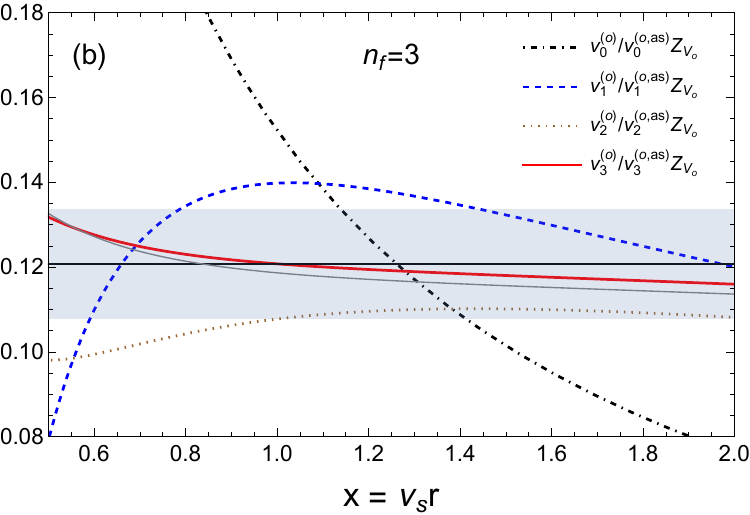}
\end{minipage}
\caption{\it Determination of $Z_{V_o}$ using $v^{(o)}_n/v_n^{(o,as)}Z_{V_o}$ as a function of $x=\nu_s r$ and for different values of $n$ in the $\MS$ scheme. The gray continuous line is  $v^{(o)}_3/v_3^{(o,as)}Z_{V_o}$ without the ultrasoft logarithmically related term. The black horizontal line is our final prediction and the blue band our final error estimate. {\bf (a)} are determinations with $n_f=0$ and {\bf (b)} with $n_f=3$.} 
\label{fig:ZVo}
\end{figure}

The discussion goes in complete parallel to the evaluation of the $Z_{V_s}$ carried out in the previous section. The prediction for $Z_{V_o}$  using $v^{(o)}_n/v_n^{(o,as)}Z_{V_o}$ as a function of $x=\nu_s r$ for different values of $n$ can be found in Fig. \ref{fig:ZVo}. The error analysis is also identical to the one of the singlet static potential case. Nevertheless, the relative magnitude of the error estimates coming from the different methods is different. Whereas in the singlet static potential, the scale variation was much bigger than the difference between the different orders. Now the scale dependence is slightly smaller than the difference between $v^{(o)}_3/v_3^{(o,as)}Z_{V_o}$ and $v^{(o)}_2/v_2^{(o,as)}Z_{V_o}$ at $x=1$. Therefore, the latter gives the bulk of the error. Our final numbers for $Z_{V_o}$ read
\bea
\label{ZVonf0}
\left.Z_{V_o}\right|_{n_f=0}&=&0.136(8)
\,,
\\
\label{ZVonf3}
\left.Z_{V_o}\right|_{n_f=3}&=&0.121(13)
\,.
\eea

We now turn to $Z_A$. Unlike $Z_{V_o}$, it has been obtained in two different ways for $n_f=0$. It was first determined in Refs. \cite{Bauer:2011ws,Bali:2013pla} for $n_f=0$ using the perturbative expansion of the energy of a static source in the adjoint representation in the lattice (Wilson action) scheme. The most up-to-date determination reads \cite{Bali:2013qla}
\be
\label{ZAnf0latt}
\left.Z_A\right|_{n_f=0}=-1.373(92)
\,.
\ee

\begin{figure}[htb] 
\begin{minipage}[b]{.49\linewidth}
  \centering\includegraphics[width=84mm]{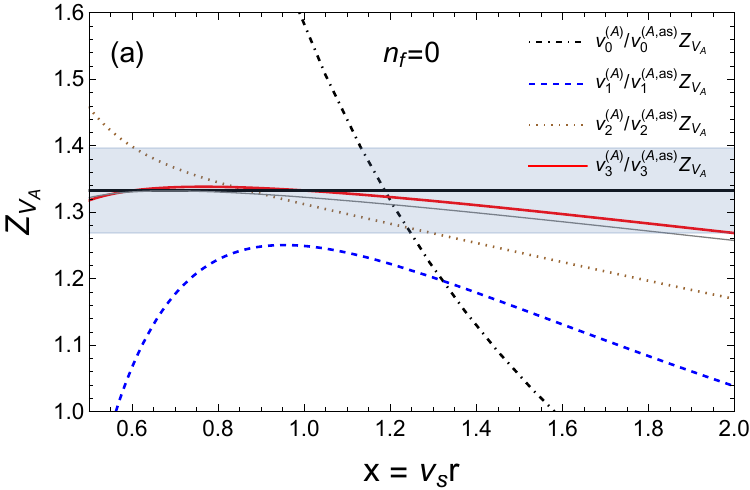}
  \end{minipage}
\begin{minipage}[b]{.49\linewidth}
  \centering\includegraphics[width=78mm]{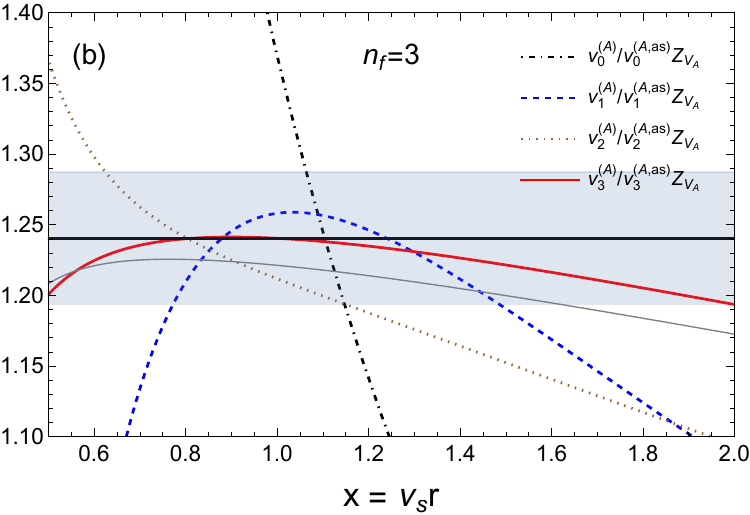}
\end{minipage}
\caption{\it Determination of $Z_{V_A}$ using $v^{(A)}_n/v_n^{(A,as)}Z_{V_A}$ as a function of $x=\nu_s r $ and for different values of $n$ in the $\MS$ scheme. The gray continuous line is  $v^{(A)}_3/v_3^{(A,as)}Z_{V_A}$ without the ultrasoft logarithmically related term. The black horizontal line is our final prediction and the blue band our final error estimate. {\bf (a)} are determinations with $n_f=0$ and {\bf (b)} with $n_f=3$.} 
\label{fig:ZA}
\end{figure}

An alternative determination of $Z_A$ follows by combining Eq. (\ref{ZmZVoZA}) and Eq. (\ref{ZmZVs}). This implies that one can use the perturbative series expansion of $V_o-V_s$. For this quantity the pole mass renormalon cancels, and we can directly obtain $Z_A$ from the perturbative expansion (in the $\MS$ scheme) of $V_A=V_o-V_s$. We can then proceed in the same way as  we did for $Z_{V_s}$ and $Z_{V_o}$. We obtain (note that $Z_{V_A}=-Z_A$)
\bea
\label{Zanf0}
\left.Z_A\right|_{n_f=0}&=&-1.332(64) \ ,
\\
\label{Zanf3}
\left.Z_A\right|_{n_f=3}&=&-1.240(47) \ ,
\eea
and in Fig. \ref{fig:ZA} we draw the prediction for $Z_{V_A}$  using $v^{(A)}_n/v_n^{(A,as)}Z_{V_A}$ as a function of $x=\nu_s r$ for different values of $n$.

For $n_f=0$ we have two independent determinations of $Z_A$ with different systematics. Therefore, we can combine them in quadrature. We obtain
\be
Z_A=-1.346(53)
\,.
\ee

Whereas this determination has the advantage of treating the perturbative series of the singlet and octet in a correlated way, improving on the error, it does not profit from the fact that the bulk of the result is coming from the singlet static potential (the octet potential contribution is $1/N_c^2$ suppressed), which is an object that can be obtained from several sources, and, therefore, the combined result has an smaller error. Indeed, if we take the quantity $Z_{V_A}=2Z_m+Z_{V_o}$, we obtain
\bea
\label{Zanf0bis}
\left.Z_A\right|_{n_f=0}&=& -1.343(36)\ ,
\\
\label{Zanf3bis}
\left.Z_A\right|_{n_f=3}&=& -1.224(43) \,,
\eea
where we have treated $Z_m$ and $Z_{V_o}$ as two independent quantities and combined the error quadratically. 

We see that the errors in Eqs. (\ref{Zanf0bis}) and (\ref{Zanf3bis}) are smaller than in Eqs. (\ref{Zanf0}) and (\ref{Zanf3}). Therefore, we take those as our final numbers. For the $n_f=0$ case, one may also consider combining Eq. (\ref{Zanf0bis}) and Eq. (\ref{ZAnf0latt}). For this case, however, they can not be considered as independent as Eq. (\ref{ZAnf0latt}) is obtained in Ref. \cite{Bali:2013qla} and the other is an average of determinations that includes one from Ref. \cite{Bali:2013qla}. Therefore, they suffer from the same systematics, and we refrain from trying to combine these two numbers. 

\subsection{Determination of subleading renormalons of the octet static potential}
\label{Sec:ZFmain}

The perturbative expansion of the octet static potential is known with the same accuracy than for the singlet static potential. Therefore, one may ask oneself whether it is also possible to obtain a determination of the subleading renormalon of the static octet potential (which is the leading renormalon of the associated force) in a similar way as it was done in Ref. \cite{Ayala:2020odx}. Nevertheless, as we have already discussed before, there is a potential problem. In the singlet potential case, the ${\cal O}(r^2)$ renormalon associated ambiguity is completely correlated with the ultraviolet renormalon of the ultrasoft contribution proportional to $V_A^2$. $V_A$ is known to be 1 with NLL accuracy. For the hybrid static energies, there are two other possible terms that contribute to ${\cal O}(r^2)$. One is proportional to $V_B^2$ and the other to $V_C$ (see Eqs. (1) and (2) of Ref. \cite{Pineda:2011db} for the operators of the effective theory). In perturbation theory, these terms are zero because there are no scales to regulate them (unlike the term proportional to $V_A^2$, which scales with the ultrasoft scale $\Delta V$). Nevertheless, they may produce effects on the asymptotic behavior of the perturbative expansion of the octet potential. The reason is that these two extra terms can have an infrared and ultraviolet renormalon. The ultraviolet ones would cancel with the corresponding infrared ones of the octet potential.  Therefore, they will affect the asymptotic behavior of the ${\cal O}(r^2)$-associated renormalon of the octet potential. On top of that, the values of $V_B$ and $V_C$ are only known to leading order in $\alpha$. Therefore, we refrain from trying to obtain the normalization of each of these subleading renormalons, nor of incorporating these effects in our determinations of the gluelump masses. The effect of these renormalons would be to add a terminant proportional to $r^2$ (times a smooth function in $\alpha$), which would mix with genuine nonperturbative/ultrasoft effects. The most we will do is explore whether the perturbative expansion is sensitive to these renormalons, and give a first estimate of what we call $Z'^{eff}_{V_o}$, the effective normalization of the subleading renormalon of the octet potential associated with these ${\cal O}(r^2)$ renormalons.

The best way to quantify the asymptotic behaviour of the perturbative series is by
performing its Borel transform:
\be
B[rV_{o,\pert}]\equiv\sum_{n=0}^{\infty}\frac{rv_n^{(o)}}{n!}\left(\frac{4\pi}{\beta_0}u\right)^{\!n}\,.
\ee
The Borel transform will have a singularity due to the dimension $d=3$ non-local condensate at $u=d/2=3/2$. We use the renormalon proportional to $V_A^2$ as a template. A more complicated renormalon structure (which would change the $1/n$ corrections) will not change the qualitative picture, and it will go into the error budget. Therefore, we approximate the subleading renormalon to be
\be
\label{eq:borel}
B[rV_{o,\pert}]
\dot=
Z'^{eff}_{V_o}(r \nu)^3 \, \frac{1}{(1-2u/d)^{1+db}}\left[1+b_1\left(1-\frac{2u}{d}\right)+\cdots\right]\,,
\ee
where $b_1=ds_1$. This singularity produces the following asymptotic behavior
\begin{align}
\label{pn}
r(V_{o,n}-V_{o,n}^{(\rm as)}) &\stackrel{n\rightarrow\infty}{=} Z'^{eff}_{V_o}(r \nu)^3\,
\left(
\frac{\beta_0}{2\pi d}\right)^{\!n}
\frac{\Gamma(n+1+db)}{\Gamma(1+db)}
\left\{
1+\frac{db}{n+db}\,b_1
+
\order\left(\frac{1}{n^2}\right)
\right\}
\,.
\end{align}

As we did in Ref. \cite{Ayala:2020odx}, we work with the force
\be
F_o(r) \equiv \frac{d}{dr}V_o(r) \sim \sum_{n=0}^{\infty}f_{o,n}\alpha^{n+1}
\,,
\ee
as this quantity automatically elliminate the ${\cal O}(r^0)$ leading renormalon. This maximizes the signal associated with these renormalons.  Asymptotics is then given by
\begin{equation}
r^2f_{o,n} \stackrel{n\rightarrow\infty}{=} Z^{eff}_{F_o} x^3\left( \frac{\beta_0}{6\pi }\right)^{n}
\frac{\Gamma(n+1+3b)}{\Gamma(1+3b)}
\left\{
1+\frac{3b}{n+3b}\,b_1
+
\mathcal{O}\left(\frac{1}{n^2}\right)
\right\}
\,.
\end{equation}
where $Z^{eff}_{F_o}=2Z'^{eff}_{V_o}$.

\begin{figure}[htb] 
\begin{minipage}[b]{.49\linewidth}
  \centering\includegraphics[width=84mm]{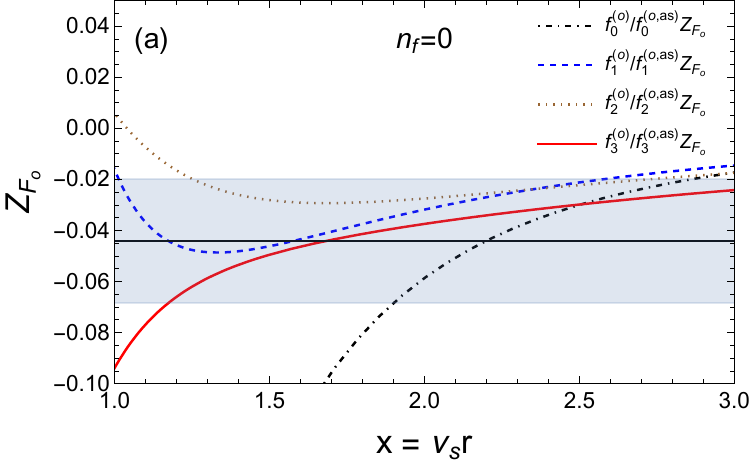}
  \end{minipage}
\begin{minipage}[b]{.49\linewidth}
  \centering\includegraphics[width=78mm]{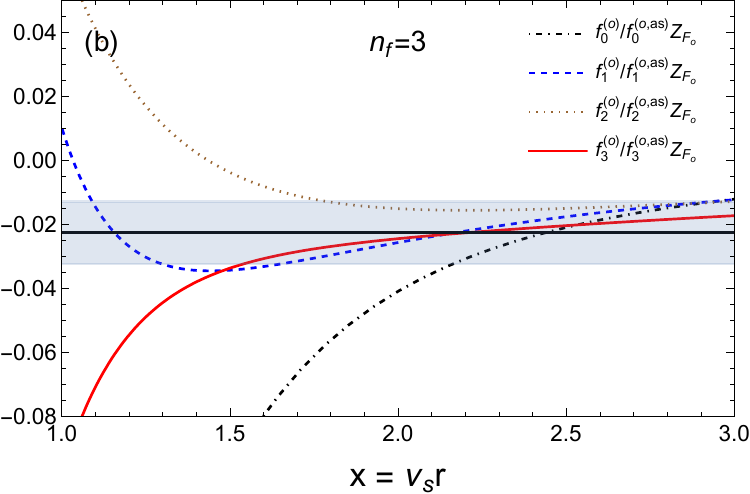}
\end{minipage}
\caption{\it Determination of $Z^{eff}_{F_o}$ using $f^{(o)}_n/f_n^{(o,as)}Z^{eff}_{F_o}$ as a function of $x=\nu_s r$ and for different values of $n$ in the $\MS$ scheme. {\bf (a)} are determinations with $n_f=0$ and {\bf (b)} with $n_f=3$.} 
\label{fig:ZFO}
\end{figure}

In Fig. \ref{fig:ZFO}, we plot the ratio of $f^{(o)}_n/f_n^{(o,as)}Z^{eff}_{F_o}$. The plot shows sensitivity to the $O(r^2)$ renormalon but the normalization is quite compatible with zero. Here, we follow the same procedure for determining the normalization of the force as in Ref. \cite{Ayala:2020odx}. We take $\nu_{us}=\nu_s$. The central value is chosen at the minimal sensitivity scale $x_{cv}$, and the associated error on $x$ is given by $x=x_{cv}/\sqrt{2}$ and $x=\sqrt{2} x_{cv}$. We obtain: $x^{F_o}_{cv}(n_f=0)=1.68$, $x^{F_o}_{cv}(n_f=3)=2.21$, $x^{F_A}_{cv}(n_f=0)=1.25$, and $x^{F_A}_{cv}(n_f=3)=1.47$ (the scale of minimal sensitivity for $F_o$ is set using one order less, since the last order does not have a minimal scale unless one goes to extremely small values of $x$). The error associated to ultrasoft effects is estimated by dropping the logarithmic-associated term. In the equations below, we list the sensitivity of the normalization to different ways to estimate the error for $n_f=0$ and $n_f=3$, and our final error:
\bea
\nn
\left.Z^{eff}_{F_o}\right|_{n_f=0}&=&
-0.044^{+0.001}_{-0.023}(\Delta x)+0.015({\rm {N}^2LO})+0.008( \mathcal{O}(1/n))+0.000(\rm {us})=-0.044(24) \ ,
\\
\label{ZFonf0}
\\
\nn
\left.Z^{eff}_{F_o}\right|_{n_f=3}&=&
-0.023^{+0.006}_{-0.009}(\Delta x)+0.007({\rm {N}^2LO})+0.003(\mathcal{O}(1/n))+0.002(\rm {us})=-0.023(10) 
\label{ZFonf3}
\ .
\eea

\begin{figure}[htb] 
\begin{minipage}[b]{.49\linewidth}
  \centering\includegraphics[width=84mm]{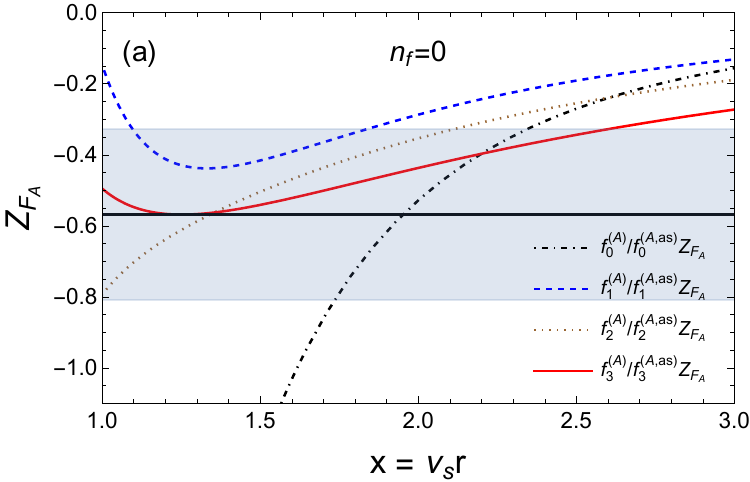}
  \end{minipage}
\begin{minipage}[b]{.49\linewidth}
  \centering\includegraphics[width=78mm]{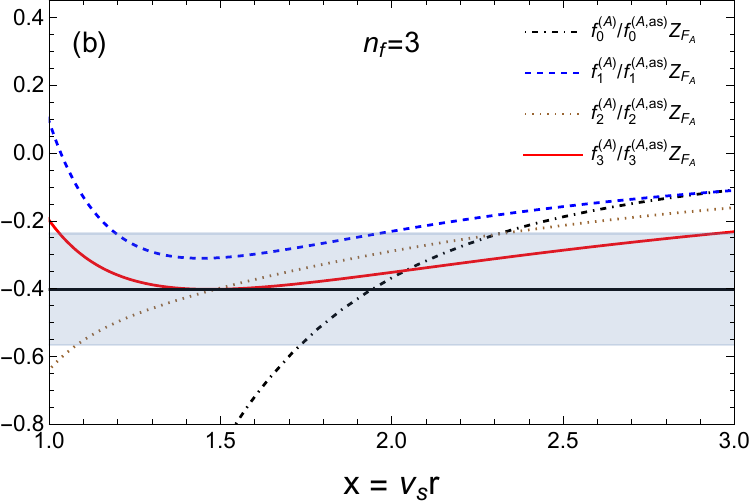}
\end{minipage}
\caption{\it Determination of $Z^{eff}_{F_A}$ using $f^{(A)}_n/f_n^{(A,as)}Z^{eff}_{F_A}$ as a function of $x=\nu_s r$ and for different values of $n$ in the $\MS$ scheme. {\bf (a)} are determinations with $n_f=0$ and {\bf (b)} with $n_f=3$.} 
\label{fig:ZFA}
\end{figure}

For completeness, we also do a similar analysis considering the force associated with $V_A$. We show the results in Fig. \ref{fig:ZFA}, and in the equations below\bea
\nn
\left.Z^{eff}_{F_A}\right|_{n_f=0}&=&-0.57^{+0.08}_{+0.21}(\Delta x)-0.04({\rm {N}^2LO})+0.11( \mathcal{O}(1/n))-0.02(\rm {us})=-0.57(24) \ ,
\\
\label{ZFanf0}
\\
\nn
\left.Z^{eff}_{F_A}\right|_{n_f=3}&=&-0.40^{+0.06}_{+0.16}(\Delta x)+0.00({\rm {N}^2LO})+0.06(\mathcal{O}(1/n))-0.00(\rm {us})=-0.40(16) \ .
\\
\label{ZFanf3}
\eea

\section{Determination of $\Lambda_{B}^{\PV}$ from the lattice gluelump energy}
\label{Sec:LambdaLatt}

\begin{figure}[htb]
\begin{center}
\includegraphics[width=0.814\textwidth]{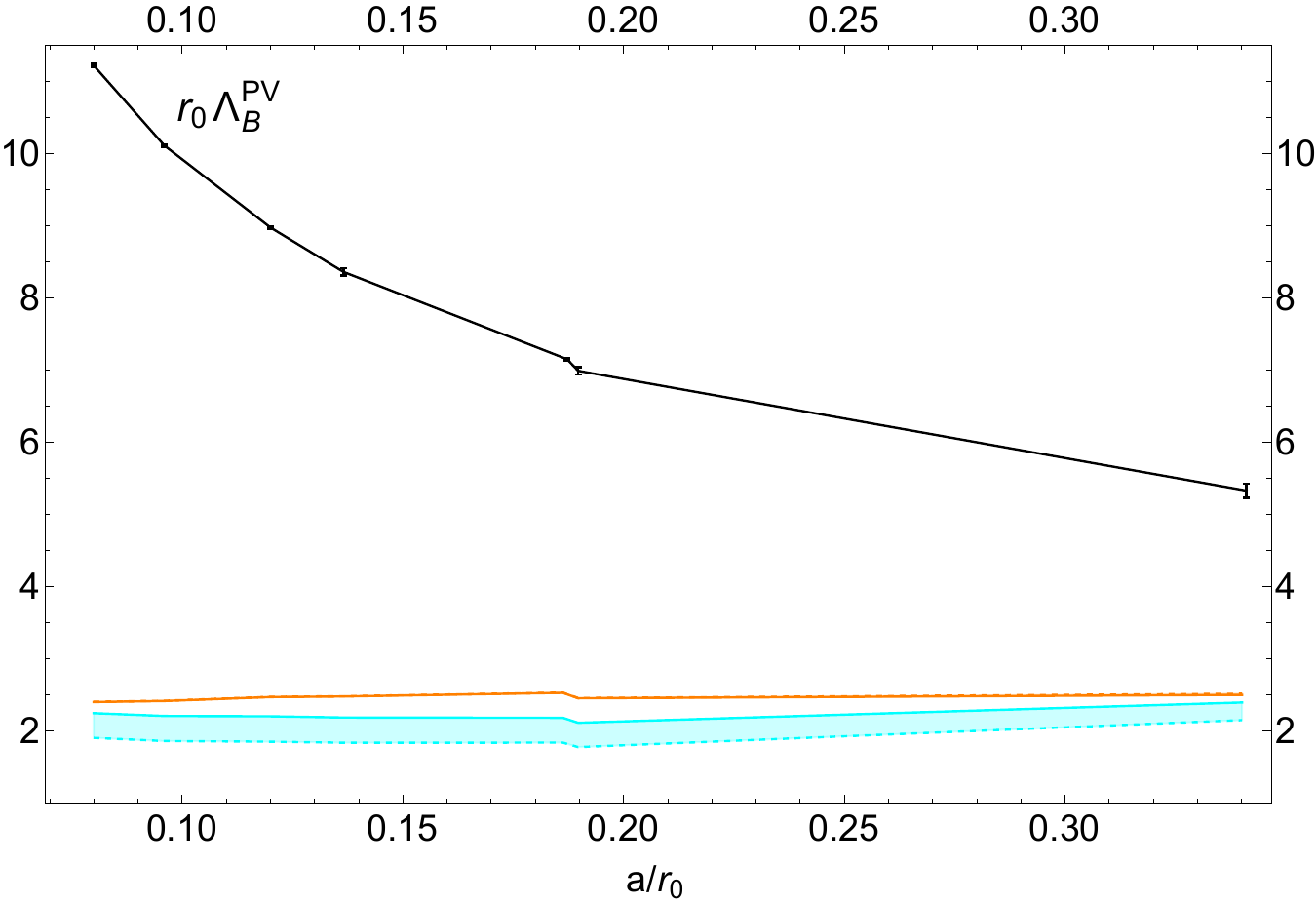}
\end{center}
\caption{The black points correspond to $\Lambda_B^L(a)$: the Montecarlo lattice data (including errors) of Refs. \cite{Foster:1998wu} and \cite{Herr:2023xwg}. The continuous lines are drawn to guide the eye. The other lines correspond to Eq. (\ref{LambdabarPV}) truncated at different orders in the hyperasymptotic expansion: (a) $\Lambda_B^L(a)-\delta m_{A}^{(P)}(1/a)$ (cyan band), (b) $\Lambda_B^L(a)-\delta m_{A}^{(P)}(1/a)-\frac{1}{a}\Omega_A$ (orange band). The bands of each order are generated by the difference of the prediction produced by the smallest positive (continuous line) or negative (dashed line) possible values of $c$ that yield integer values for $N_P$.}
\label{Fig:barLambdaLatt1}
\end{figure}
\begin{figure}[htb]
\begin{center}
\includegraphics[width=0.82\textwidth]{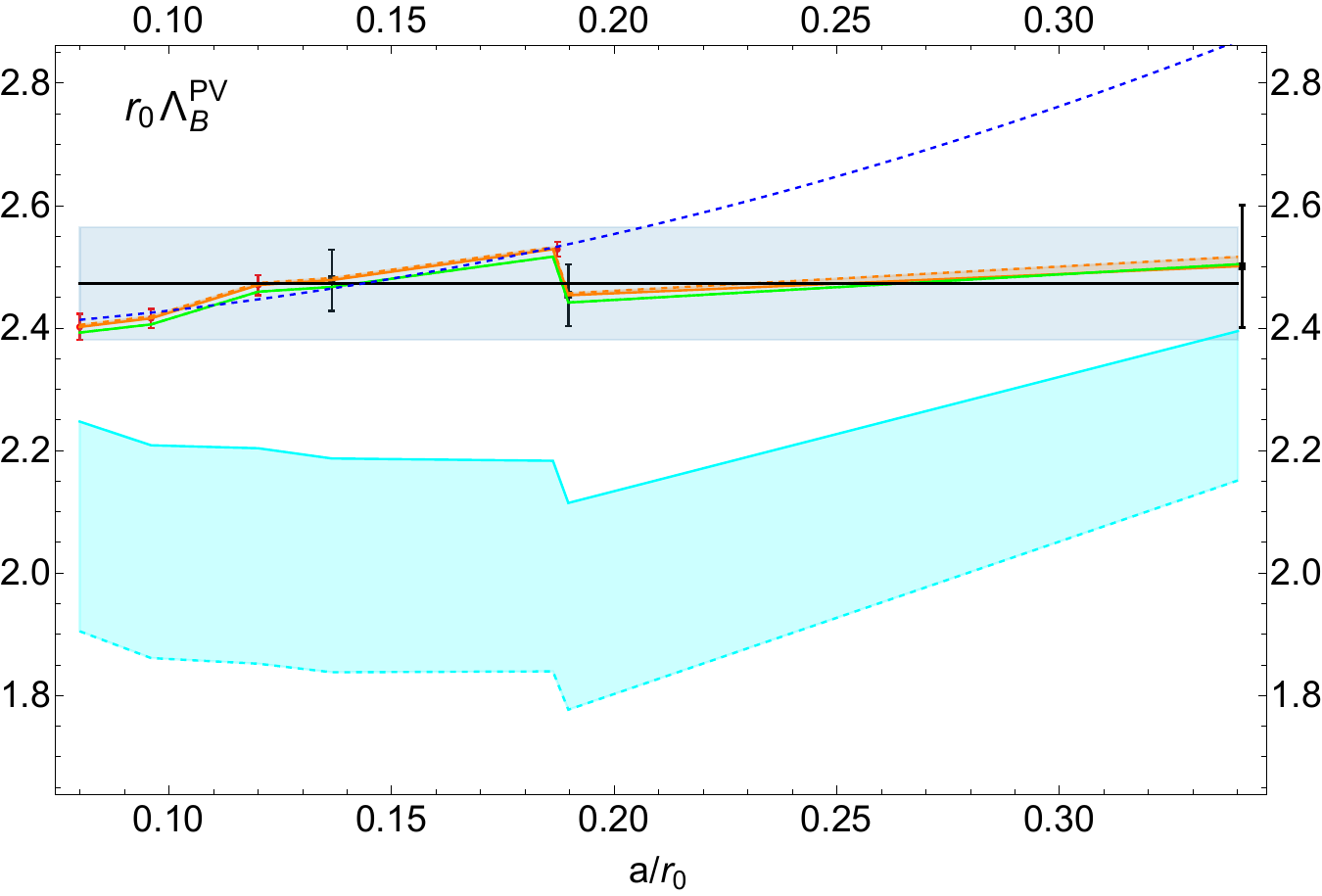}
\end{center}
\caption{As in the previous figure but in a smaller range. We plot Eq. (\ref{LambdabarPV}) truncated at different orders in the hyperasymptotic expansion: (a) $\Lambda_B^L(a)-\delta m_{A}^{(P)}(1/a)$ (cyan band), (b) $\Lambda_B^L(a)-\delta m_{A}^{(P)}(1/a)-\frac{1}{a}\Omega_A$ (orange band), (c) $\Lambda_B^L(a)-\delta m_{A}^{(P)}(1/a)-\frac{1}{a}\Omega_A-\sum_{N_P+1}^{13}\frac{1}{a}[c_{A,n}-c_{A,n}^{\rm (as)}]\alpha_L^{n+1}$ (green line). 
The bands are generated by the difference of the prediction produced by the smallest positive (continuous line) or negative (dashed line) possible values of $c$ that yield integer values for $N_P$.
In (b) we include the error of the lattice points in the continuous line and distinguish between the data set \cite{Foster:1998wu} (black points) and \cite{Herr:2023xwg} (red points). The horizontal black line is our final fit and the horizontal blue band our error estimate. The dashed blue line is a fit to (b) = $\Lambda_B+Ka^2$ using only the \cite{Herr:2023xwg} data points.}
\label{Fig:barLambdaLatt2}
\end{figure}

We will consider the lightest gluelump, $H=B$. 
We will use the data \cite{Foster:1998wu} used in \cite{Bali:2003jq} (see Table 3 in that reference) and the data \cite{Herr:2023xwg} (see second row in Table 9) for $\Lambda_B^L(a)$. All this data is quenched. Effects associated with light fermions should change the results up to $\sim$ 10\%. We neglect possible correlations between the lattice data points. This effect will not be much important in comparison with other sources of error. 

The formula we use for the determination of $\Lambda_{B}^{\PV}$  is the following (see Eqs. (\ref{LambdaHL}) and (\ref{deltamAPV}))
\be
\label{LambdabarPV}
\Lambda_{B}^{\PV}=\Lambda_B^L(a)-\delta m_{A}^{(P)}(1/a)-\frac{1}{a}\Omega_A(1/a;a)-\sum_{N_P+1}^{N'=3N_P}\frac{1}{a}[c_{A,n}-c_{A,n}^{\rm (as)}]\alpha_L^{n+1}(a)+{\cal O}(a^2)
\,.
\ee
Note that the left-hand side and the right-hand side of this equation do not depend on $a$, the lattice spacing (or should not within errors). We plot the right-hand side  of Eq. (\ref{LambdabarPV}) with increasing level of accuracy in the hyperasymptotic expansion in Figs. \ref{Fig:barLambdaLatt1} and \ref{Fig:barLambdaLatt2}. We can observe a convergent pattern for the subtraction of the different terms that appear in the hyperasymptotic expansion. Indeed, it is easy to see that the bulk of the value of $\Lambda_B^L(a)$ is associated with perturbation theory. After reaching superasymptotic approximation (cyan band; (0,$N_P$) in the hyperasymptotic counting), the result is relatively flat, reflecting the independence in the lattice spacing. For the lattice spacing we are considering (and in the lattice Wilson action strong coupling scheme), we need to take $N_P\sim 6$. 
The inclusion of the leading terminant (orange band; (1,0) in the hyperasymptotic counting) makes the result even more $a$-independent and the dependence in $c$ nearly vanishes. To go beyond  (1,0) precision in the hyperasymptotic counting, we need to add the term 
$$
\sum_{N_P+1}^{N'=3N_P}\frac{1}{a}[c_{A,n}-c_{A,n}^{\rm (as)}]\alpha_L^{n+1}.
$$
 As we already discussed in Sec. \ref{Sec:gluelumps1}, such inclusion is delicate.  Each individual term is extremely large. It requires a very precise cancellation (with exponential accuracy) between the exact and asymptotic coefficients. We saw in Sec. \ref{Sec:gluelumps1} that for $Z_A=-1.343$, such cancellation deteriorates for $n > 13$. We also saw that if we move away from the value $Z_A=-1.343$ within the error range in Eq. (\ref{Zanf0bis}), the deterioration of the cancellation takes place at even smaller values of $n$.\footnote{This suggests that it should be possible to get a more accurate determination of $Z_A$. In particular, we believe that there is room for improvement here by a dedicated/correlated study of the infinite volume extrapolation of the exact perturbative coefficients, the coefficients of the beta function in the Wilson action scheme, and the value of $Z_A$. We will not explore such possibility in this paper, though.} Therefore, the most we will do is to add the term 
\be
\label{incomplete}
\sum_{N_P+1}^{13}\frac{1}{a}[c_{A,n}-c_{A,n}^{\rm (as)}]\alpha_L^{n+1}
\ee
to reach (1,13) precision in the hyperasymptotic counting and to use the difference with the (1,0) result as a partial estimate of this subleading term of the hyperasymptotic expansion. We display the result as the green line in Fig. \ref{Fig:barLambdaLatt2}. The correction to the value of $\Lambda_{B}^{\PV}$ is tiny ($\sim 0.011$), changing little with respect the (1,0) result, which we will take as our default value. This result is not very sensitive to the error of $Z_A$. We find the error to be rather small ($\sim 0.008$). It is also not very sensitive to evaluations taking $c$ positive or negative. We find a very small error ($\sim 0.0034$). On the other hand, the result is relatively sensitive to the expression used for $\Omega$. In the lattice scheme, the convergence of the sum in Eq. (\ref{eq:OmegaV}) is slow. For the central value, we use Eq. (\ref{eq:OmegaV}) with three terms in the sum (this needs estimates of the higher order coefficients of the beta function in the lattice scheme, which we take from Ref. \cite{Ayala:2019hkn}). To estimate the error we do the computation using Eq. (\ref{eq:OmegaV}) with only two terms in the sum. The error due to the truncation of the expansion in $\Omega$ is around $\sim 0.025$, which we will add to the error budget. 

This (1,0) (or (1,13)) precision in the hyperasymptotic counting allows us to start seeing finer details of the lattice data. We highlight them in Fig. \ref{Fig:barLambdaLatt2}. The newer data \cite{Herr:2023xwg} have smaller errors than the previous data set \cite{Foster:1998wu}, and it seems to show a slope that the older data do not show. This seems to indicate that the ${\cal O}(a^2)$ effects are different between the two data sets, being much bigger in \cite{Herr:2023xwg} than in \cite{Foster:1998wu}. Therefore, such difference cannot be attributed to the missing terms in Eq. (\ref{incomplete}), since they are equal for both lattice data sets, and small (according to the magnitude of Eq. (\ref{incomplete})). Overall, our theoretical determination, at present, does not reach enough precision to quantify the possible tension between these data sets. Therefore, such ${\cal O}(a^2)$ effects go into the error budget. In order to avoid bias in our theoretical determination, we determine the central value by fitting the combined data sets to a constant with (1,0) hyperasymptotic precision. We obtain $2.47(2)r_0^{-1}$. The resulting $\chi^2=7.8$ is quite large, but the central value is almost identical to the central value one obtains if one only fits the data of \cite{Foster:1998wu} to a constant, resulting in the same central value: $2.47(3)r_0^{-1}$ but with a small $\chi^2=0.1$ in this case. We also consider a quadratic fit: $\Lambda_{B}^{\PV}+ca^2$ to the data set of \cite{Herr:2023xwg}. We obtain $2.39(2)r_0^{-1}$. The resulting $\chi^2=1.34$ is quite reasonable.\footnote{The analysis of \cite{Herr:2023xwg} using the RS scheme still shows a relatively large $\chi^2$, even after including these ${\cal O}(a^2)$ terms. In our view, this indicates that the hyperasymptotic expansion accounts for the perturbative contribution to $\Lambda_B^{\PV}$ more accurately.}  If, instead, we fit the new lattice data to $\Lambda_{B}^{\PV}+ca$, we obtain  $2.31(3)r_0^{-1}$ with a reasonable $\chi^2=0.73$. This indicates that the precision of the lattice data is not enough to distinguish the functional form of the remaining $a$ dependence. Nevertheless, we rule out this possibility based on theoretical arguments. 

In order to make the final estimate of the error, we combine the statistical error of the fit ($\sim 0.021$), the error associated with the approximate knowledge of the terminant, i.e., of $Z_A$ ($\sim 0.008$) and of $\Omega$ ($\sim 0.025$), and the difference ($\sim 0.86$) between the fits to $\Lambda_{B}^{\PV}$ and to $\Lambda_{B}^{\PV}+ca^2$. By far the latest gives the bulk of the error, which we take as the measure of subleading terms in the hyperasymptotic expansion (including nonperturbative effects). 
Overall, we obtain
\be
\label{fit1}
\Lambda_{B}^{\PV}=2.47(9) r_0^{-1}\simeq 1\; \rm GeV
\,.
\ee

Note that, in the above error analyses, there is no error associated with $\Lambda_{QCD}$, since we know  exactly the value of $\beta$, and, therefore, of $\alpha_L(a)=3/(2\pi\beta)$. We have also considered the error associated with the Necco-Sommer formula is negligible. 

\section{Determination of $\Lambda_{B}^{\PV}$ from the static hybrid energy}

It is also possible to determine $\Lambda_{B}^{\PV}$ from the short-distance behavior of the energies of the $\Pi_u^-$ or $\Sigma_u^-$ static hybrids. The former is a better choice because it has been determined with smaller errors. These static hybrid energies are sensitive to the renormalon of the pole mass. In their determinations using lattice simulations, this reflects in the fact that the result has a $1/a$ dependence.  We would like our determination not to be affected for possible inaccuracies in our knowledge of this renormalon. Therefore, we consider the energy difference $E_{\Pi_u}-E_{\Sigma_g^+}$  for which we consider its hyperasymptotic expansion, written in Eq. (\ref{EPiuSigmag}), and isolate $\Lambda_B^{\PV}$ from there:
\bea
\nn
\Lambda_B^{\PV}&=&(E_{\Pi_u}(r)-E_{\Sigma_g^+}(r))
-V_{A,P}(r)-\frac{1}{r}\Omega_{V_A}- \delta V_A^{\rm RG}(r) 
\\
&&
-\sum_{n=N_P+1}^{3N_P/N_{max}} (V^{(A)}_n-V_n^{(A,\rm as)}) \al_{\MS}^{n+1}(\nu_s)
-\delta E^{(2)\PV}_{A,us}(r;\nu_{us})
+o( r^2)
\label{Method2}
\eea

The lattice error of $E_{\Sigma_g^+}$ is much smaller than the error of $E_{\Pi_u}$. Therefore, it can be neglected, and we will only consider the lattice error of $E_{\Pi_u}$.

$E_{\Pi_u}-E_{\Sigma_g^+}$ also has the advantage that one can do the continuum limit in the lattice. In practice, this is achieved by dedicated studies such as the one one may find in Ref. \cite{Bali:2003jq}. Alternatively, one can work at the very finest lattices and see that points at different lattice spacings lie on top of each other within errors. This is the attitude we take with the data set from \cite{Schlosser:2021wnr}. In the first case, the continuum error is included in the error of the lattice points. In the second, the error is smaller than the error of the individual points and will, accordingly, be neglected.
\begin{center}
\begin{figure}
\begin{center}
\includegraphics[width=0.814\textwidth]{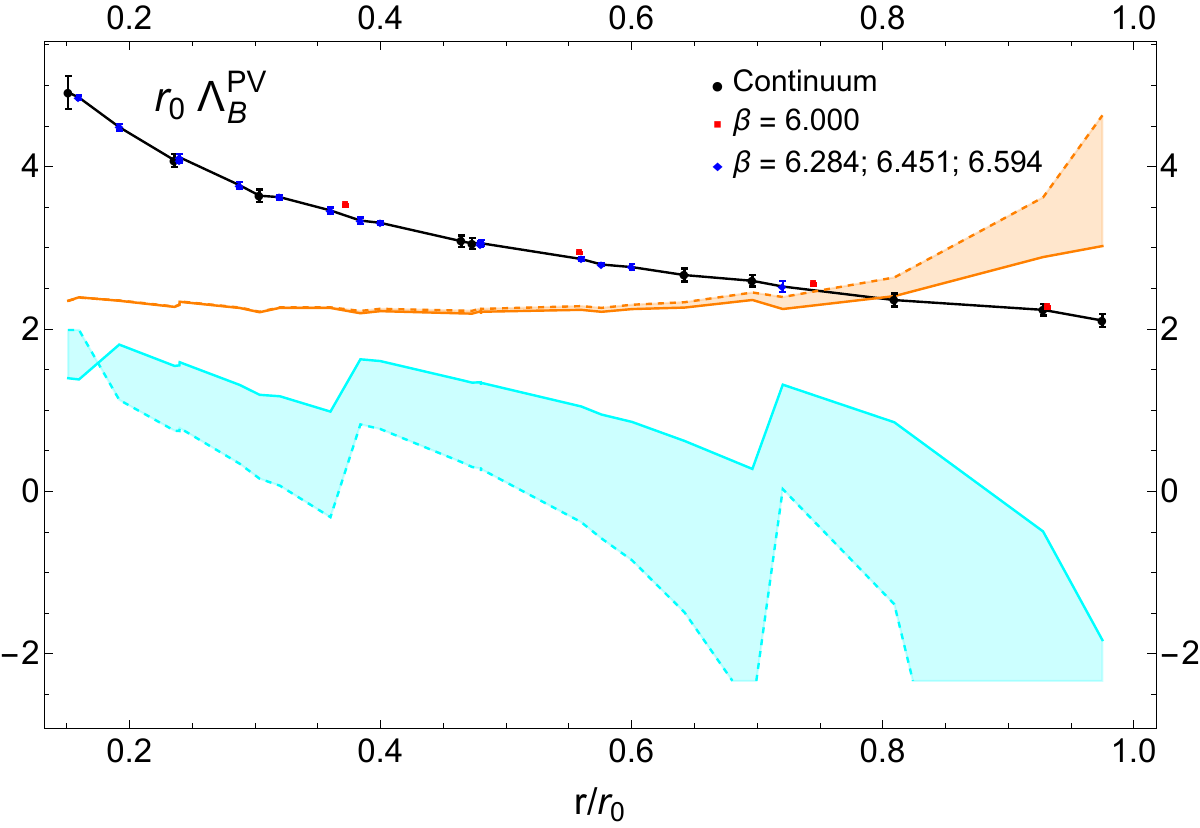}
\end{center}
\caption{The Montecarlo lattice data points correspond to $E^L_{\Pi_u}(r)-E^L_{\Sigma_g^+(r)}$: \cite{Bali:2003jq} (black points) and \cite{Schlosser:2021wnr} (blue points, except for those at $\beta=6$, which we draw in red). The data set from \cite{Bali:2003jq} is already in the continuum limit. The continuous lines are drawn to guide the eye. The other lines correspond to Eq. (\ref{Method2}) truncated at different orders in the hyperasymptotic expansion:
(a) $E^L_{\Pi_u}(r)-E^L_{\Sigma_g^+}(r)-V_{A,P}^{(P)}(r)$ (cyan band), (b) $E^L_{\Pi_u}(r)-E^L_{\Sigma_g^+}(r)
-V_{A,P}^{(P)}(r)-\frac{1}{r}\Omega_{V_A}$ (orange band). For each difference, the bands are generated by the difference of the prediction produced by the smallest positive (continuous line) or negative (dashed line) possible values of $c$ that yield integer values for $N_P$ up to the two most-to-the-left points. For extra details see the main text. }
\label{Fig:LambdaB21}
\end{figure}
\end{center} 
We plot the right hand side of Eq. (\ref{Method2}) with an increasing degree of accuracy in the hyperasymptotic expansion in Fig. \ref{Fig:LambdaB21}. The continuous black line just corresponds to the lattice result for $E^L_{\Pi_u}(r)-E^L_{\Sigma_g^+(r)}$. The other lines correspond to Eq. (\ref{Method2}) truncated at different orders in the hyperasymptotic expansion: (a) $E^L_{\Pi_u}(r)-E^L_{\Sigma_g^+}(r)-V_{A,P}^{(P)}(r)$ (cyan band; (0,$N_P$)), (b) $E^L_{\Pi_u}(r)-E^L_{\Sigma_g^+}(r)-V_{A,P}^{(P)}(r)-\frac{1}{r}\Omega_{V_A}$ (orange band; (1,0)). For each order, the bands are generated by the difference of the prediction produced by the smallest positive (continuous line) or negative (dashed line) possible values of $c$ that yield integer values for $N_P$.
We can observe a convergent pattern for the subtraction of the different terms that appear in the hyperasymptotic expansion. After reaching superasymptotic approximation, (0, $N_P$), the result is somewhat flatter. This starts reflecting the to-be independence in $1/r$ of Eq. (\ref{Method2}) (up to ${\cal O}(r^2)$ effects). On the other hand, the result is strongly dependent on $c$. This is a reflection that, unlike in the lattice scheme considered in the previous section, the magnitude of the different terms of the perturbative series in the $\MS$ scheme is much bigger. Note also that, for the two most-to-the-left points in the blue band, we do not have enough terms in the perturbative expansion, since $N_P > N_{max}$ for the smallest $c$ negative. Therefore, for these two points, we take the variation associated with $c$ ($N_P$) to be between the two smallest possible positive values of $c$, but still keeping the central value to be the $c$ with the smallest possible value.

 \begin{center}
\begin{figure}
\begin{center}
\includegraphics[width=0.82\textwidth]{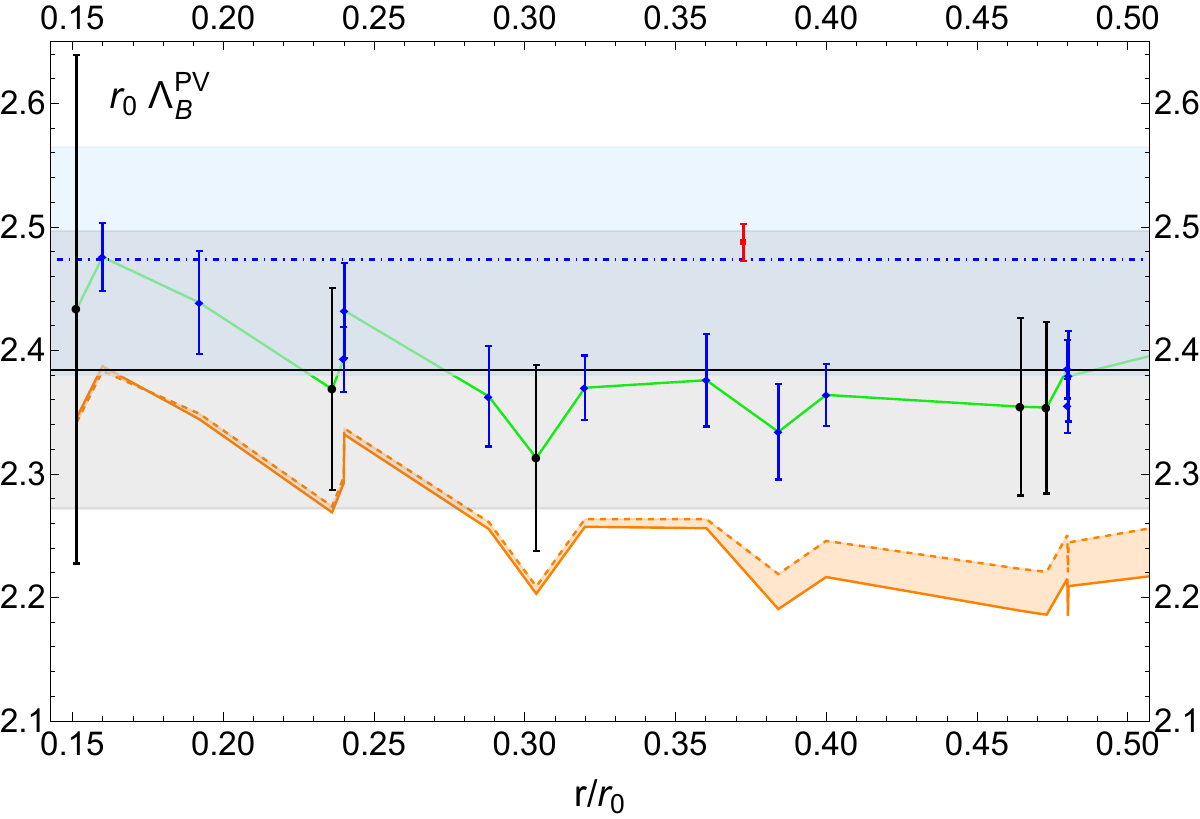}
\end{center}
\caption{As in the previous figure but in a smaller range. 
We plot Eq. (\ref{Method2}) truncated at different orders in the hyperasymptotic expansion: (b) $E^L_{\Pi_u}(r)-E^L_{\Sigma_g^+}(r)-V_{A,P}^{(P)}(r)-\frac{1}{r}\Omega_{V_A}$ (orange band), (c) $E^L_{\Pi_u}(r)-E^L_{\Sigma_g^+}(r)-V_{A,P}^{(P)}(r)-\frac{1}{r}\Omega_{V_A}-\delta V_{A,\rm RG}(r)+\sum_{N_P+1}^{N'=N_{max}}\frac{1}{a}[c_{A,n}-c_{A,n}^{\rm (as)}]\alpha_{\MS}^{n+1}-\delta E^{(2)\PV}_{o,us}(r;\nu_{us})$ (green line), where $\delta E^{(2)\PV}_{o,us}(r;\nu_{us})$ is computed at leading order in perturbation theory, and we have included the error of the lattice points in the middle of the green line (black points are those coming from Ref. \cite{Bali:2003jq}; blue points from Ref. \cite{Schlosser:2021wnr} with $\beta=6.284, 6.451$ and $6.594$, and the red point comes from the same reference with $\beta=6$). The orange band is generated by the difference of the prediction produced by the smallest positive (continuous line) or negative (dashed line) possible values of $c$ that yield integer values for $N_P$ up to the two most-to-the-left points. For extra details see the main text. The black horizontal line is the fit to the green line and the grey band is our final error (see Eq. (\ref{eq:method2})). For comparison, we have also displayed the fit from Sec. \ref{Sec:LambdaLatt} displayed in Eq. (\ref{fit1}) (the blue band and dot-dashed horizontal line).}
\label{Fig:LambdaB2}
\end{figure}
\end{center}
 The strong dependence in $c$ basically disappears once the leading terminant is included (orange band; (1,0)). Including this term makes the result to be $r$-independent in a good approximation, and up to relatively large distances. This level of precision allows us to start seeing finer details of the lattice data. We highlight them in Fig. \ref{Fig:LambdaB2}. After zooming in on the plot, we see that $E^L_{\Pi_u}(r)-E^L_{\Sigma_g^+}(r)-V_{A,P}^{(P)}(r)-\frac{1}{r}\Omega_{V_A}$ (the orange band) still has some residual $r$ dependence, especially for the points most to the left. One possible explanation is that the points most to the left could be affected by possible discretization effects since $r=2a$ for them. We also try to improve the prediction by adding more terms to the hyperasymptotic expansion. We do so by adding $\delta V_A^{\rm RG}$, the resummation of the ultrasoft logarithms with $\nu_{us}=\sqrt{2}/r_0$ ($\alpha(\sqrt{2}/r_0)=0.557$), and 
\be
\sum_{N_P+1}^{N'=N_{max}}\frac{1}{a}[c_{A,n}-c_{A,n}^{\rm (as)}]\alpha_{\MS}^{n+1}
\,.
\ee
We find that the effect of this term is not numerically important. Indeed, except for the seven most-to-the-right points, for which $N_P=1$ (so $3N_P+1=N_{max}+1=4$), we do not have enough terms in the perturbative expansion to reach the next renormalon, and for the two most-to-the-left points this term is zero.  On the other hand, the effect of the resummation of logarithms is important and accounts for most of the difference between the orange and green line. It is also worth noticing that including the resummation of logarithms makes the result slightly flatter. One may wonder that the resummation of logarithms is formally an ${\cal O}(r^2)$ effect, but, on the other hand, it is logarithmically enhanced. Finally, we also include an estimate of $\delta E^{(2)\PV}_{o,us}(r;\nu_{us})$. We use Eq. (\ref{dEUSpert}) in order to make our prediction as weak-coupling based as possible (in a strict weak-coupling counting, this would be NNNLL accuracy). We then take the fit to Eq. (\ref{Method2}) using Eq. (\ref{dEUSpert}) as the central value. The fit to the lattice data yields $\Lambda_B^{\PV}=2.384(8)\; r_0^{-1}$ with a quite reasonable $\chi^2=1.29$.
 Eq. (\ref{dEUSpert}) contributes little to this result. If we simply subtract it, the difference is $\sim 0.012$ with a similar $\chi^2$. 

{\bf Terminant}. We now estimate the error associated with our incomplete knowledge of the terminant. The situation is somewhat reversed compared with the analysis in the previous section. In that section, the analysis was made in the lattice scheme, whereas now we are working in the $\MS$ scheme.  In the lattice scheme, the error associated with $Z_A$, the normalization of the renormalon, was very small; here it is the opposite. If we take $Z_A=-1.307$, the difference is $\sim -0.07$ but the agreement with the data significantly deteriorates ($\chi^2=3.5$). On the other hand, if we take $Z_A=-1.379$, the difference is $\sim 0.07$ and the consistency with the lattice data improves, reflecting in a smaller $\chi^2=0.86$. We will add the maximum of this difference (being quite symmetric) to the error budget. Note also that this last result is in better agreement with the result obtained in the previous section. We now consider the error associated with the truncation of the expression we use for $\Omega$. In the $\MS$ scheme, the convergence is much better and we can safely work in the strict weak coupling limit (see Eq. (\ref{eq:OmegaV2})). Truncating to one order less, the difference with the central value is $\sim 0.007$, and it can be safely neglected. The variation associated with $c$ is even smaller $\sim 0.004$ and also neglected. 

$\boldsymbol{\Lambda_{\MS}}$. We now consider the variation associated with the incomplete knowledge of $\Lambda_{\MS}=0.602(48)\;r_0^{-1}$ \cite{Capitani:1998mq}. If we take $\Lambda_{\MS}=0.650\;r_0^{-1}$ we obtain $\Lambda_B^{\PV} \sim 2.46 \; r_0^{-1}$ with a good $\chi^2=1.44$. For $\Lambda_{\MS}=0.554\;r_0^{-1}$, the agreement deteriorates ($\chi^2=4.68$) and $ \Lambda_B^{\PV}\sim 2.34 \; r_0^{-1}$. We add the maximum of these differences to the error budget.

{\bf Higher order terms in the hyperasymptotic expansion}\\
We now consider errors more specific to higher-order terms in the hyperasymptotic expansion. 

{\bf $\boldsymbol{{\cal O}(r^2)}$ terms}. We first analyze how reliable is using Eq. (\ref{dEUSpert}) for the ${\cal O}(r^2)$ ultrasoft correction. As we discussed before, dropping this term changes little the result ($\sim 0.01$). Instead, we can add an arbitrary $A r^2$ term to the fit. The difference with the central value is $\sim 0.045$ with $\chi^2=0.83$. We will add this difference to the error budget. 

$\boldsymbol{\nu_{us}}$. The value we have taken for $\nu_{us}$ is quite small: $\nu_{us}=\sqrt{2}/r_0$. One reason we have taken this value is to ensure that when we do the variation of $\nu_s$, we always have that $\nu_{us} < \nu_s$ for the whole range $r$ that we consider. We have also observed that such a small value of $\nu_{us}$ yields a better fit than if, for instance, we set $\nu_{us}=2/r_0$. In this last case, the agreement with lattice data deteriorates, which reflects in a bigger value of $\chi^2=2.86$. We also observe that the difference with the determination obtained in the previous section increases (we obtain $\Lambda_B^{\PV}=2.31\; r_0^{-1}$). Fits adding an arbitrary $A r^2$ are, however, rather stables to changes of $\nu_{us}$. If we take $\nu_{us}=2/r_0$, the difference with the central value is $\sim 0.01$. Therefore, we do not add an extra error from this analysis.  

$\boldsymbol{\nu_{s}}$. With respect to the variation of $\nu_s$, we consider the range $\nu_s \in (1/\sqrt{2}r, 2r)$. Choosing the upper limit of the range: $\nu_s=2r$ produces a bad fit ($\chi^2=3.87$) with $\Lambda_B^{\PV}=2.28\; r_0^{-1}$. Therefore, we disregard it. The lower limit of the range: $\nu_s=1/\sqrt{2}r$ produces a good fit ($\chi^2=1.25$) with $\Lambda_B^{\PV}=2.42\; r_0^{-1}$. The difference with the central value is smaller than considering an extra quadratic term in the fit. Therefore, we do not include it in the final error budget to avoid double counting. 

In order to make the final estimate of the error, we combine the statistical error of the fit ($\sim 0.008$), the error associated with $\Lambda_{\MS}$ ($\sim 0.07$), the error associated with the approximate knowledge of the terminant, i.e. of $Z_A$ ($\sim 0.07$), and the difference ($\sim 0.04$) between the fits to $\Lambda_{B}^{\PV}$ and to $\Lambda_{B}^{\PV}+Ar^2$. 
Overall, our final prediction for $\Lambda_B^{\PV}$ reads
\be
\label{eq:method2}
\Lambda_B^{\PV}=2.38(11)\;r_0^{-1}\,.
\ee

In Fig. \ref{Fig:LambdaB2}, we compare this result (horizontal black line and grey band) with the green line, which corresponds to the theoretical expression Eq. (\ref{Method2}) using Eq. (\ref{dEUSpert}), and where we have made explicit the lattice points (with their errors). The new \cite{Schlosser:2021wnr} and old \cite{Bali:2003jq} data are perfectly compatible with each other within errors. There is only (an expected) difference for $\beta=6.0$, reflecting that, for such $\beta$, one has not yet obtained the continuum limit of $E^L_{\Pi_u}(r)-E^L_{\Sigma_g^+}(r)$ for the precision we have achieved. This means that for this $\beta$, $O(a^2)$ effects are clearly distinguishable. This is the reason we disregard lattice data points for $\beta=6.0$ (only one point in the range of $r$ we consider). On the other hand, finer lattices are perfectly consistent with each other and with the continuum lattice data obtained in Ref. \cite{Bali:2003jq}. On the theoretical side, we would expect a large cancellation of lattice artifacts in the energy difference  $E^L_{\Pi_u}(r)-E^L_{\Sigma_g^+}(r)$. It would be interesting to study where this remaining ${\cal O}(a^2)$ dependence comes from. Note also that we would expect to have ${\cal O}(r^2)$ effects. Nevertheless, we can not really quantify this possible effect because of errors. A dedicated analysis is postponed. 

Let us now make some comparisons with the analysis made in Sec. \ref{Sec:LambdaLatt}. 
The error of the lattice points is typically bigger now than in the previous case. Also, the analysis now scans over larger distances than in the previous case. The observable we are considering in this section is sensitive to the ultrasoft scale, which complicates the analysis. On the other hand, as a matter of principle, it is an object for which we can do the continuum limit. For the specific observables we are looking at, this is not much of an advantage, since the $1/a$ running in the previous section plays the same role as the $1/r$ running in this section and it is well under control within the hyperasymptotic expansion. Overall, the precision we have obtained in this section is slightly worse than the one obtained in the previous section, but both predictions are perfectly compatible within errors.

\begin{center}
\begin{figure}
\begin{center}
\includegraphics[width=0.814\textwidth]{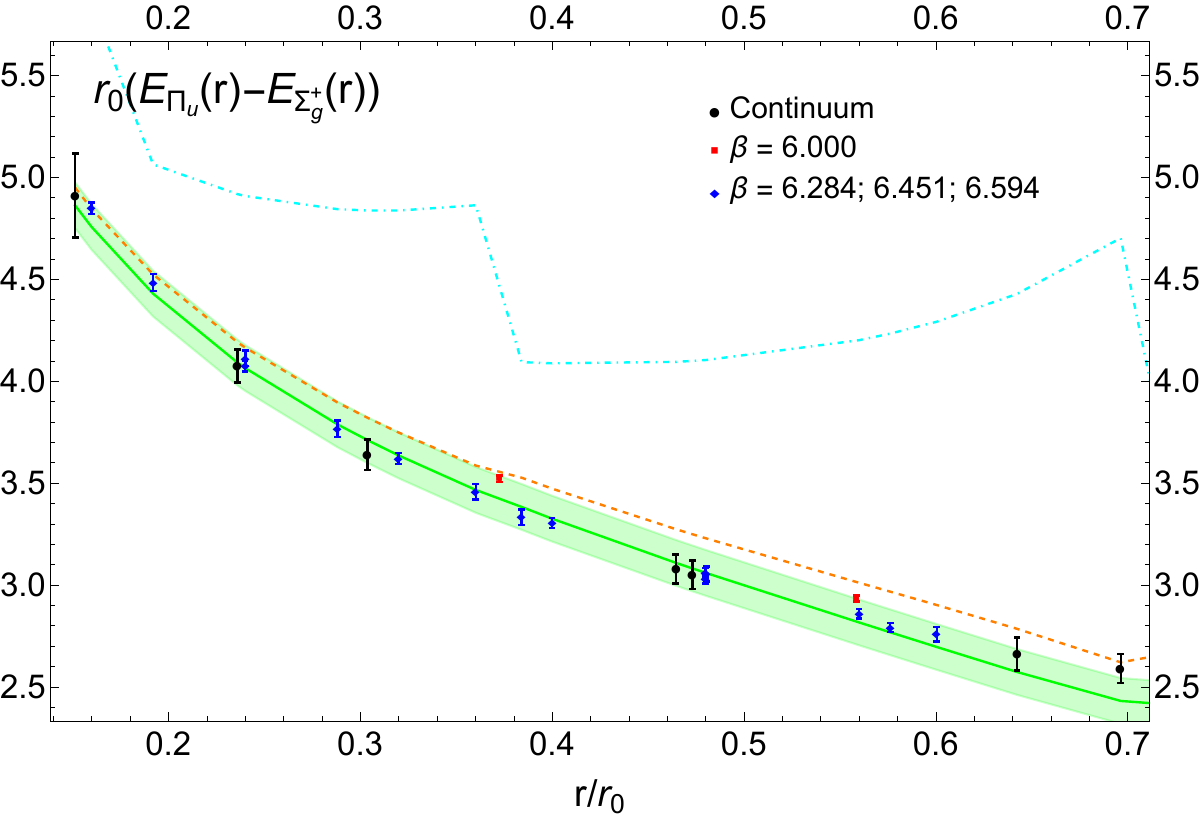}
\end{center}
\caption{The Montecarlo lattice data points correspond to $E^L_{\Pi_u}(r)-E^L_{\Sigma_g^+(r)}$: \cite{Bali:2003jq} (black points) and \cite{Schlosser:2021wnr} (blue points, except for those at $\beta=6$, which we draw in red). The dot-dashed blue line corresponds to $V_{A,P}(r)+\Lambda_B^{\PV}$ (with $c$ positive), the dashed orange line to $V_{A,P}(r)+\Lambda_B^{\PV}+\frac{1}{r}\Omega_{V_A}$, and the green line to 
$V_{A,P}(r)+\Lambda_B^{\PV}+\frac{1}{r}\Omega_{V_A}+ \delta V_A^{\rm RG}(r) 
+\sum_{n=N_P+1}^{N_{max}} (V^{(A)}_n-V_n^{(A,\rm as)}) \al^{n+1}(1/r)-\delta E^{(2)\PV}_{o,us}(r;\nu_{us})$.
Finally, the green band is generated by the error of $\Lambda_B^{\PV}$ in Eq. (\ref{eq:method2}).}
\label{Fig:Pot}
\end{figure}
\end{center}

As a final plot, we show in Fig. \ref{Fig:Pot} how the hyperasymptotic expansion approximates the energy difference $E_{\Pi_u}-E_{\Sigma_g^+}$. We find rather reasonable agreement up to scales well beyond $r=r_0/2$. The breakdown occurs at around $r \sim 0.7 r_0$.

\subsection{${\cal O}(r^2)$}

We now study the energy difference between the hybrids $\Sigma_u^-$ and $\Pi_u$: $E_{\Sigma_u^-}-E_{\Pi_u}$. This quantity is completely free of perturbative effects. Based on EFT arguments, we expect this energy difference to scale as $r^2$:
\be
E_{\Sigma_u^-}-E_{\Pi_u}=A_{\Sigma_u^-\Pi_u}r^2
\,.
\ee

\begin{center}
\begin{figure}
\begin{center}
\includegraphics[width=0.814\textwidth]{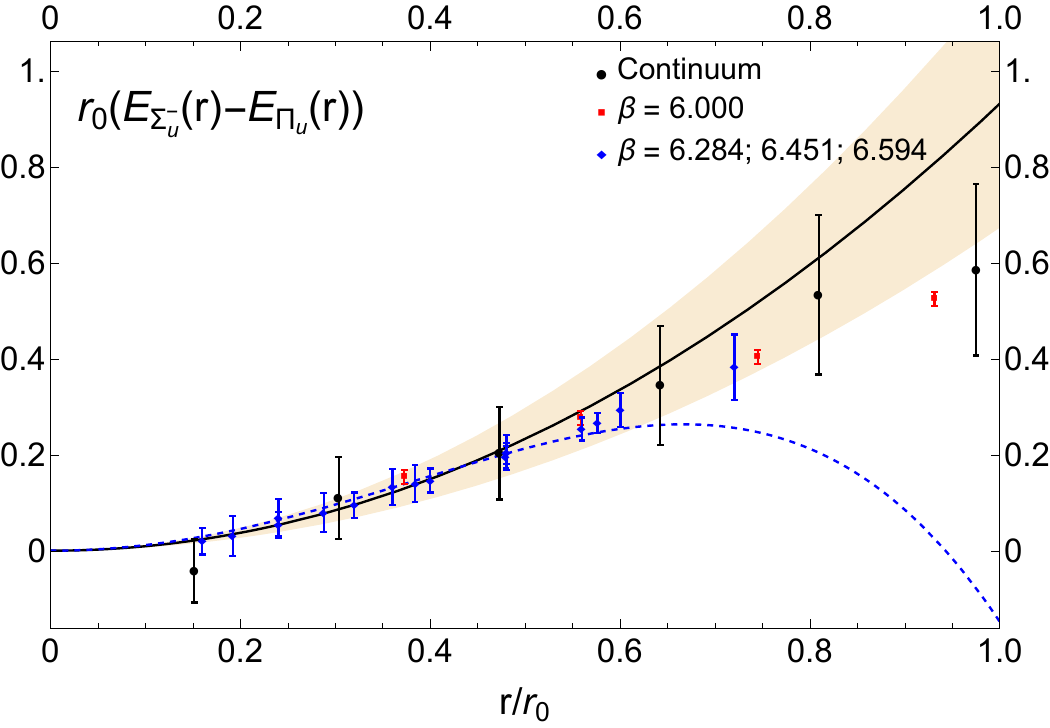}
\end{center}
\caption{We plot $E^L_{\Sigma_u^-}-E^L_{\Pi_u}$. The black data points are taken from \cite{Bali:2003jq}, where the  continuum limit result was obtained. The red and blue points are the data obtained in Ref. \cite{Herr:2023xwg} for different lattice spacings. The black line is the quadratic fit  and the blue dashed line the quadratic plus quartic fit. The band reflects the error of our determination. }
\label{Fig:r2}
\end{figure}
\end{center} 

We show a plot of this quantity in Fig. \ref{Fig:r2}, as well as the result of a fit till $r_0/2$. The result of the quadratic fit yields
\be
A_{\Sigma_u^-\Pi_u}=0.93(26)\;r_0^{-3}
\,.
\ee
The final error is the combination in quadrature of the systematic and statistical error. The first is completely dominant and has been determined by considering the difference between the value obtained from the pure quadratic fit and a fit also including a quartic $\sim r^4$ term. We show the prediction for both fits in the figure. The number we obtain in this paper is very similar to the number obtained in Ref. \cite{Bali:2003jq} but with smaller errors. 

It is worth noticing that for this energy difference, $E^L_{\Sigma_u^-}-E^L_{\Pi_u}$, ${\cal O}(a^2)$ lattice artefacts seems to largely cancel so that the $\beta=6$ simulation lies on top of the other simulations. Therefore, we do not exclude it from our fits (though, since we are talking of a single point in the fit region, its effect in the fit is tiny). 

As a final remark, Fig. \ref{Fig:r2}, together with Fig. \ref{Fig:Pot}, can be used as a test of the range where one can apply the multipole expansion (and a weak-coupling EFT based analysis). Reasonable agreement is reached up to scales well beyond $r=r_0/2$. The breakdown occurs at around $r \sim 0.7 r_0$.

\section{Conclusions}
We give the most up-to-date determinations of the normalization of the leading renormalons of the pole mass, the singlet static potential, the octet static potential, and the gluelump energy. They read
\be
Z^{\MS}_m=-Z^{\MS}_{V_s}/2(n_f=0)=0.604(17)
\,, \qquad Z^{\MS}_m=-Z^{\MS}_{V_s}/2(n_f=3)=0.551(20)
\,,
\ee
\be
Z^{\MS}_{V_o}(n_f=0)= 0.136(8) 
\,,
\qquad Z^{\MS}_{V_o}(n_f=3)=0.121(13)
\,,
\ee
\be
Z^{\MS}_{A}(n_f=0)=-1.343(36) 
\,,\qquad Z^{\MS}_{A}(n_f=3)=-1.224(43)
\,.
\ee
These numbers are consistent, albeit more precise, than older values obtained in Refs. \cite{Pineda:2001zq,Bali:2003jq,Bauer:2011ws,Bali:2013pla,Bali:2013qla,Ayala:2014yxa,Beneke:2016cbu}.

We obtain two independent RG invariant and renormalization scale independent determinations of the energy of the ground state gluelump in the principal value summation scheme. The first method directly used the gluelump mass computed in the lattice. Fitting this object to its hyperasymptotic expansion, we obtain $\Lambda_{B}^{\PV}=2.47(9)r_0^{-1}$. This result is remarkably stable to different lattice spacings. The second method considers the small distance limit of the energy difference of the 
$\Pi_u(r)$ and $\Sigma_g^+$ states in the static limit. This method yields $\Lambda_{B}^{\PV}=2.38(11)r_0^{-1}$. The dominant error for each method comes from different systematics. Therefore, it makes sense to combine them in quadrature. This produce our final result:
\be 
\label{LambdaFinal}
\Lambda_{B}^{\PV}=2.44(7)r_0^{-1}.
\ee
This number comes from the combination of two different determinations. Each of them corresponds to an evaluation at different scales and, above all, with very different renormalization schemes for the strong coupling (lattice and $\MS$ scheme). The fact that both determinations agree within errors is a very strong check of our result. In comparison, previous determinations \cite{Bali:2003jq,Herr:2023xwg} were computed in the RS scheme and had bigger errors. If we compare with those results, by changing the scheme, we find these numbers are slighlty smaller than our determination, in particular the result of Ref. \cite{Herr:2023xwg}. Nevertheless, they are perfectly compatible with our result within errors. It should be mentioned that the change to the RS scheme is made using the strong coupling constant in the $\MS$ scheme, as this is how previous results were computed and the RS scheme result is dependent on that. 

The fact that we can give an absolute number for $\Lambda_{B}^{\PV}$ allows us to give absolute numbers for the masses of all gluelumps. One only has to sum our value to the values given in Table 2 of Ref. \cite{Herr:2023xwg} to have scheme/scale independent numbers for the absolute values of the gluelump masses. 

The fact that we get good agreement for the ground state hybrid potential up to relatively long distances opens the possibility of using this result in the determination of the spectrum and properties of some heavy quarkonium hybrid states. The values we have obtained of the gluelump masses can be directly put in first principle computations of the hybrid spectrum when solving the Schroedinger equations. The heavy quark masses that appear in such computations are not free parameters but the values obtained from first principle computations using the same summation scheme, and similarly for the potentials.  This possibility will be considered in future work together with the incorporation of light dynamical quarks.

Finally, it is interesting to see that $\Lambda_{B}^{\PV} \sim 2 \bar \Lambda^{\PV}$ obtained in \cite{Ayala:2019hkn}. This is consistent with the expectation that the scaling of $\Lambda_{B}$ goes with $C_A/2$ and of $\bar \Lambda$ with $C_F/2$. Within perturbation theory, $\bar \Lambda$ would be proportional, within a good approximation, to $C_F=(N_c^2-1)/(2N_c)$ and $\Lambda_B$ to $C_A/2$. The question is whether the numbers multiplying the casimir coefficients are approximately universal in the nonperturbative regime. This issue was raised in the context of static potentials between sources in different colour representations in Ref. \cite{Ambjorn:1984dp}
and studied in the lattice in Ref. \cite{Bali:2000un}. In this paper, we have seen that our results are consistent with casimir scaling, even though they are not static potentials (but closely related to those). We do not assign error estimates, as our aim here is just to numerically check that such hypothesis is sensible in the context of this work. 

\medskip
   
\noindent
{\bf Acknowledgments.}\\
This work was supported in part by FONDECYT (Chile) Grants No. 1240329, by the Spanish Ministry of Science and Innovation 
 (PID2020-112965GB-I00 and PID2023-146142NB-I00), and by the Departament de Recerca i Universities from Generalitat de Catalunya to the Grup de Recerca 00649 (Codi: 2021 SGR 00649). This project has received funding from the European Union's Horizon 2020 research and innovation programme under grant agreement No 824093. IFAE is partially funded by the CERCA program of the Generalitat de Catalunya.


\end{document}